\documentclass[apj]{emulateapj}

\shorttitle{Overtone and double-mode RR Lyrae stars in M3}
\shortauthors{Jurcsik et al.}

\begin{document}

\title{Overtone and multi-mode RR Lyrae stars in the globular cluster  M3}

\author{J. Jurcsik\altaffilmark{1}, P. Smitola\altaffilmark{1},  G. Hajdu\altaffilmark{2,3}, \'A. S\'odor\altaffilmark{1}, J. Nuspl\altaffilmark{1}, K. Kolenberg\altaffilmark{4,5},  G. F\H ur\'esz\altaffilmark{6}, A. Mo\'or\altaffilmark{1}, E. Kun\altaffilmark{7,8}, \\ A. P\'al\altaffilmark{1}, J. Bakos\altaffilmark{1}, J. Kelemen\altaffilmark{1}, T. Kov\'acs\altaffilmark{1}, L. Kriskovics\altaffilmark{1}, K. S\'arneczky\altaffilmark{1}, T. Szalai\altaffilmark{9}, A. Szing\altaffilmark{10} K. Vida\altaffilmark{1}} 

\affil{\altaffilmark{1}Konkoly Observatory of the Hungarian Academy of Sciences, H-1525 Budapest PO Box 67, Hungary}
\affil{\altaffilmark{2}Instituto de Astrof\'{i}sica, Pontificia Universidad Cat\'olica de Chile, Av. Vicu\~na Mackenna 4860, 782-0436 Macul, Santiago, Chile}
\affil{\altaffilmark{3}Millennium Institute of Astrophysics, Av. Vicu\~na Mackenna 4860, 782-0436 Macul, Santiago, Chile}
\affil{\altaffilmark{4}Institute of Astronomy, KU Leuven, Celestijnenlaan 200D, 3001 Heverlee, Belgium}
\affil{\altaffilmark{5}Harvard-Smithsonian Center for Astrophysics, 60 Garden Street, Cambridge MA 02138, USA}
\affil{\altaffilmark{6}MIT   Kavli Institute for Astrophysics and Space Research 77 Mass Ave 37-515, Cambridge, MA, 02139}
\affil{\altaffilmark{7}Department of Theoretical Physics, University of Szeged,  H-6720 Szeged, Tisza Lajos krt 84-86, Hungary}
\affil{\altaffilmark{8}Department of Experimental Physics and Astronomical Observatory, University of Szeged, H-6720 Szeged, D\'om t\'er 9, Hungary}
\affil{\altaffilmark{9}Department of Optics and Quantum Electronics, University of Szeged, H-6720 Szeged, D\'om t\'er 9, Hungary}
\affil{\altaffilmark{10}Baja Observatory, University of Szeged, 6500 Baja, KT: 766, Hungary}

\email{jurcsik@konkoly.hu}

\begin{abstract}
The overtone and multi-mode RR Lyrae stars in the globular cluster M3 are studied using a 200-d long, $B,V$ and $I_{\mathrm C}$ time-series photometry obtained in 2012. 70\% of the 52 overtone  variables observed show some kind of multi-periodicity (additional frequency at  ${f_{0.61}}={f_{\mathrm {1O}}}/0.61$ frequency ratio, Blazhko effect, double/multi-mode pulsation, period doubling). A signal at 0.587 frequency ratio to the fundamental-mode frequency is detected in the double-mode star, V13, which may be identified as the second radial overtone mode. If this mode-identification is correct, than V13 is the first RR Lyrae star showing triple-mode pulsation of the first three radial modes. Either the Blazhko effect or the ${f_{0.61}}$ frequency (or both of these phenomena) appear in 7 double-mode stars. The $P_{\mathrm{1O}}/P_{\mathrm{F}}$ period ratio of RRd stars showing the Blazhko effect are anomalous. A displacement of the main frequency component at the  fundamental-mode  with the value of  modulation frequency (or its half) is detected in three Blazhko RRd stars parallel with the appearance of the overtone-mode pulsation. The ${f_{0.61}}$ frequency appears in  RRc stars that lie at the blue side of the double-mode region and in RRd stars, raising the suspicion that its occurrence  may be connected to double-mode pulsation. 
The changes of the Blazhko and double-mode properties of the stars are also reviewed using the recent and archive photometric data.

\end{abstract}


\section{Introduction}\label{intsec}
Studies of cluster variables (RR Lyrae stars) date back to the end of the 19th century. The large number of the variables discovered in galactic and extragalactic globular clusters (GCs) provided a decisive support to specify the main physical properties and the evolutionary status of RR Lyrae stars. Because of the large distance  ($d>6$ kpc) and the relatively small absolute dimension (total radius, $R_t<50$~pc) of GCs, the apparent magnitudes of the cluster members reflect the distribution of their absolute magnitudes with $1-2$\% accuracy  (supposing that the foreground/intra-cluster reddening is homogeneous). The advantage that the relative values of the physical properties of GC stars can be determined more accurately than for any other stellar population makes GC studies very prosperous.

While numerous studies of fundamental-mode RR Lyrae stars (RRab) have appeared year by year in the literature, this is not the case for the overtone (RRc) variables. As RRc stars are relatively rare objects and the similarity of their light-curves to W UMa binaries makes even the target selection dubious, hence only a few field RRc stars have been the subject of detailed photometric and/or spectroscopic studies. The photometric data of GCs showed that about 30\% of their RR Lyrae populations are overtone and double-mode (RRd) pulsators \citep{cl01}. RRd stars have always been in the focus of the interest because of their astrophysical importance, and the discovery of peculiar features in the pulsations of RRc stars \citep[][and references therein]{mo14,mo15,netzel,smo15a,smo15b} has attracted a special attention to overtone RR Lyrae stars recently. Higher order or multiple pulsations together with modulation or period doubling  make these stars to unique targets of theoretical investigations. The complex and interrelated dynamical features can give strong constraints on non-linear physics governing their variability and pose a new challenge to further theoretical work, providing a test bed for improved models.

Thanks to the new data secured by space missions \citep[CoRoT, Kepler:][]{corot,sz14} and ground-based projects \citep{aip, geos}, a lot of new results were published in the recent years concerning the Blazhko effect; the amplitude and/or phase modulation of the pulsation of RR Lyrae stars. Despite these observational results and the accompanying theoretical efforts revealing the genuine origin of the phenomenon has eluded explanation yet. The modulation was detected in about 50\% of the RRab stars \citep{kbs1,kepler}, but no reliable Blazhko statistics exists for overtone variables. 

One of the most intriguing new question of overtone RR Lyrae and Cepheid stars is the recent discovery of a peculiar frequency at 0.61 frequency ratio in many of these variables \citep[][and references therein]{so10,dz12,mo15}. The interpretation of this strange signal  is hard in the recent theoretical framework because of its unknown excitation mechanism.

M3 is one of GCs with a large RR Lyrae population that was investigated in many photometric studies previously. However, even the most extended CCD observations [e.g.  \citet[][K98]{K98}, \citet[][C01]{co01},  \citet[][H05]{H05}, \citet[][B06]{be06}, and \citet[][J12]{oc}] were not long and dense enough and/or used only one filter, therefore our knowledge on the multi-mode and Blazhko properties of M3 variables is still incomplete. 
The present article gives an in-depth study of the overtone and double-mode RR Lyrae stars of M3, with full details on their modulations and multi-mode properties, as well as on the long-term changes of the light-curve variations based on an extended new multi-color time-series photometry of the cluster and on archive data.
Some interesting features of the Blazhko and double-mode properties of overtone variables using the new data set published here were already published in \cite{rrdbl}.

\begin{figure*}
\includegraphics[width=18cm]{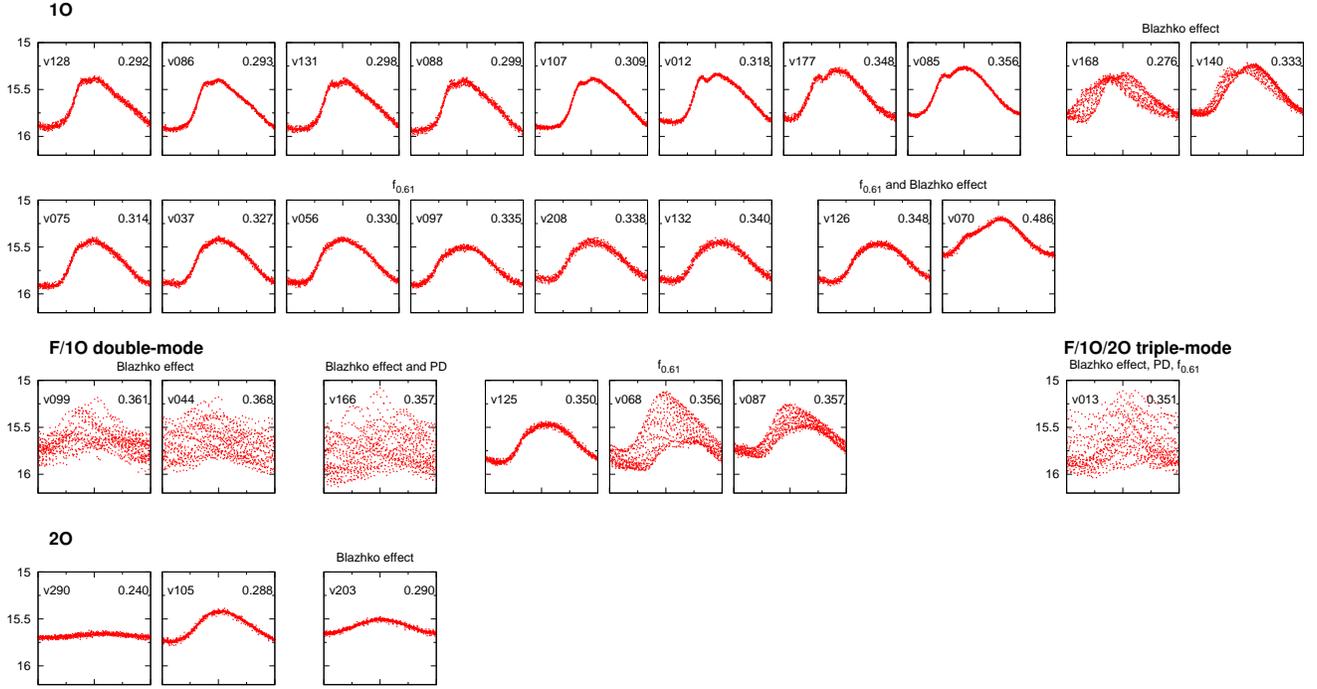}
\caption{$V$ light curves of overtone and double-mode variables in M3 phased with the overtone-mode period.  The names and periods (days) of the variables are given in the top-left- and top-right-side of each panel.\label{lcv}}
\end{figure*}

\begin{figure*}
\begin{center}
\includegraphics[width=14.7cm]{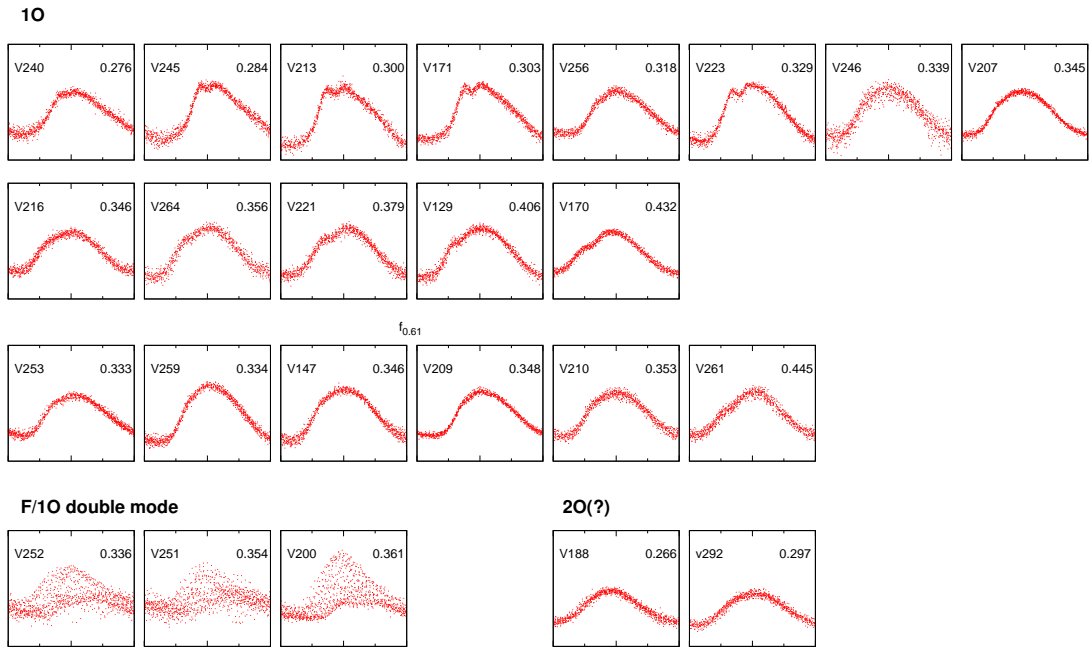}
\caption{Non-calibrated $B$-filter light curves of overtone and double-mode variables in M3 phased with the overtone-mode period. The ISIS fluxes  of the  variables with uncertain magnitudes on the reference frame  have been  transformed to magnitudes matching the mean brightness of the horizontal branch, however, the amplitudes of these light curves are uncertain.\label{lcb}}
\end{center}
\end{figure*}

\begin{table}
\begin{center}
\caption{$B$, $V$, $I_{\mathrm{C}}$ CCD observations of M3 variables published in this paper. The complete table is given in the electronic edition.\label{photmag}}
\begin{tabular}{lccc}
\hline
\hline
Star\tablenotemark{a}&HJD$-2400000$\tablenotemark{b}& Mag & Band \\
\hline
V012&55935.64720 &16.088  &B\\
V012&55937.69026 &15.583  &B\\
V012&55937.70074 &15.636  &B\\
...&  ...    &     ... &     ...\\
\hline
\end{tabular}
\tablenotetext{1}{Identification according to the catalogue of \cite{samus}}
\tablenotetext{2}{The HJDs represent the times of mid-exposure.}
\end{center}
\end{table}
\begin{table}
\begin{center}
\caption{Flux data  of M3 variables with uncertain magnitudes in the reference frame in $B$ band. The complete table is given in the electronic edition.\label{photflux}}
\begin{tabular}{lccc}
\hline
\hline
Star\tablenotemark{a}&HJD$-2400000$\tablenotemark{b}& Flux\tablenotemark{c}  \\
\hline
V129&55935.64720 & 15583.9 \\
V129&55937.69026 & 16631.5 \\
V129&55937.70074 & 15274.6 \\
...&  ...    &     ... &   \\
\hline
\end{tabular}
\tablenotetext{1}{Identification according to the catalogue of \cite{samus}}
\tablenotetext{2}{The HJDs represent the times of mid-exposure.}
\tablenotetext{3}{$B$ band flux differences of ISIS photometry \citep{isis}.}
\end{center}
\end{table}

\begin{table*}
\begin{center}
\caption{Summary of the photometric properties of the overtone-mode variables in M3.\label{rrc.dat}}
\begin{tabular}{@{\hspace{2mm}}l@{\hspace{3mm}}l@{\hspace{3mm}}c@{\hspace{3mm}}c@{\hspace{1mm}}c@{\hspace{3mm}}c@{\hspace{1mm}}c@{\hspace{3mm}}c@{\hspace{1mm}}c@{\hspace{5mm}}c@{\hspace{2mm}}c@{\hspace{2mm}}c@{\hspace{5mm}}r@{\hspace{3mm}}c}
\hline
\hline
Star\tablenotemark{a} &mode\tablenotemark{b} &period &$<B>$\tablenotemark{c} &($B$)\tablenotemark{c}&$<V>$&($V$) &$<I_{\mathrm C}>$& ($I_{\mathrm C}$)  &$A_B$\tablenotemark{d}   &$A_V$\tablenotemark{d}&$A_{I_{\mathrm C}}$\tablenotemark{d}  &\multicolumn{2}{l}{\hspace{-5mm}additional feature \tablenotemark{e}     } \\
&&d&mag&mag&mag&mag&mag&mag&mag&mag&mag&&\\
\hline
V12 &1O& 0.31786 &15.829 &15.802 &15.600 &15.584 &15.264 &15.258 &0.65 &0.51 &0.32 && \\
V37 &1O&0.32663 &15.909 &15.885 &15.664 &15.650 &15.311 &15.306 &0.64 &0.48 &0.30&$f_{0.61}$ & \\
V56 &1O&0.32960 &15.908 &15.884 &15.657 &15.643 &15.278 &15.272 &0.62 &0.47 &0.29&$f_{0.61}$ & \\
V70 &1O&0.48607 &15.691 &15.677 &15.398 &15.391 &14.963 &14.960 &0.53-0.49 &0.40-0.38 &0.25-0.24&$f_{0.61}$ &Bl \\
V75 &1O&0.31408 &15.924 &15.899 &15.686 &15.672 &15.306 &15.300 &0.63 &0.49 &0.30&$f_{0.61}$ & \\
V85 &1O &0.35581 &15.746 &15.720 &15.521 &15.505 &15.190 &15.184 &0.65 &0.51 &0.32 & &\\
V86 &1O&0.29266 &15.914 &15.886 &15.679 &15.663 &15.367 &15.361 &0.69 &0.54 &0.33 & & \\
V88 &1O &0.29875 &15.923 &15.893 &15.693 &15.675 &15.369 &15.362 &0.70 &0.54 &0.32 & &\\
V97 &1O&0.33493 &15.973 &15.955 &15.699 &15.689 &15.329 &15.325 &0.53 &0.40 &0.26 &$f_{0.61}$ &\\
V105 &2O&0.28775 &15.787 &15.776 &15.581 &15.576 &15.299 &15.297 &0.42 &0.32 &0.20 & & \\
V107 &1O&0.30903 &15.896 &15.868 &15.663 &15.646 &15.327 &15.320 &0.68 &0.52 &0.33 & &  \\
V126 &1O&0.34841 &15.935 &15.917 &15.665 &15.655 &15.284 &15.280 &0.60-0.53 &0.44-0.40 &0.30-0.26&$f_{0.61}$&Bl \\
V128 &1O& 0.29204 &15.894 &15.866 &15.657 &15.641 & --      & --      &0.69 &0.52 & --  &&  \\
V129 &1O &0.40604 &&&&&&&&&& &\\
V131 &1O&0.29769 &15.909 &15.882 &15.684 &15.668 &15.399 &15.392 &0.67 &0.52 &0.34 &&\\
V132 &1O&0.33986 &15.909 &15.892 &15.653 &15.643 &15.259 &15.247 &0.53 &0.41 &0.25 &$f_{0.61}$ &\\
V140 &1O&0.33316 &15.743 &15.722 &15.515 &15.502 &15.155 &15.150 &0.69-0.60 &0.53-0.45 &0.32-0.27 &&Bl \\
V147$^*$ &1O&0.34647 &&&&&&&&&&$f_{0.61}$ &\\                     
V168 &1O&0.27594 &15.803 &15.788 &15.588 &15.580 &15.345 &15.342 &0.65-0.46 &0.48-0.35 &0.32-0.23&&Bl \\
V170$^*$  &1O& 0.43238 &&&&&&&&&&& \\ 
V171$^*$  &1O& 0.30328 &&&&&&&&&&&  \\
V177 &1O&  0.34836 &15.763 &15.735 &15.552 &15.536 &15.263 &15.257 &0.67 &0.52 &0.33&& \\
V188$^*$  &2O:&0.26651 &&&&&&&&& && \\
V203 &2O&0.28979 &15.766 &15.764 &15.581 &15.580 &15.296 &15.296 &0.21-0.17 &0.17-0.12 &0.09-0.08 &&Bl \\
V207 &1O& 0.34530 &&&&&&&&& && \\ 
V208 &1O&0.33838 &15.923 &15.905 &15.647 &15.638 &15.266 &15.262 &0.55 &0.40 &0.26 &$f_{0.61}$ &\\
V209 &1O&0.34829 &&&&&&&&&&$f_{0.61}$ &\\
V210 &1O&0.35295 &&&&&&&&&&$f_{0.61}$ &\\
V213$^*$  &1O& 0.30001 &&&&&&&&&&&\\ 
V216 &1O& 0.34648 &&&&&&&&&&&\\ 
V221 &1O& 0.37878 &&&&&&&&&&&\\
V223  &1O& 0.32921 &&&&&&&&&&&\\ 
V240  &1O& 0.27601 &&&&&&&&&&&\\ 
V245$^*$  &1O& 0.28403 &&&&&&&&&&&\\
V246$^*$  &1O& 0.33916 &&&&&&&&&&&\\ 
V253$^*$  &1O& 0.33260 &&&&&&&&&&$f_{0.61}$&\\
V256 &1O& 0.31806 &&&&&&&&&&& \\
V259 &1O& 0.33351 &&&&&&&&&&$f_{0.61}$&\\
V261$^*$  &1O& 0.44473 &&&&&&&&&&$f_{0.61}$&\\
V264$^*$  &1O& 0.35646 &&&&&&&&&&&\\
V290 &2O& 0.24041 &15.861 &15.861 &15.679 &15.679 &15.415 &15.415 &0.05 &0.04 &0.03&&\\
V292 &2O:& 0.29654 &&&&&&&&&&& \\
\hline
\end{tabular}
\tablenotetext{1}{Asterisk denotes stars with photometry contaminated by signals of nearby variables.}
\tablenotetext{2}{1O: 1st overtone; 2O:2nd overtone.} 
\tablenotetext{3}{$<>$ and () denote magnitude and intensity averaged mean magnitudes, respectively.}
\tablenotetext{4}{Peak to peak amplitudes. Maximum and minimum amplitudes stars are given for Blazhko .}
\tablenotetext{5}{$f_{0.61}$: frequency at $f_{\mathrm{1O}}/f_{0.61}\approx 0.61-0.62$; Bl: Blazhko effect.}
\end{center}
\end{table*}

\begin{table*}
\begin{center}
\caption{Summary of the photometric properties of double/triple-mode variables in M3.\label{rrd.dat}}
\begin{tabular}{@{\hspace{0mm}}l@{\hspace{2mm}}l@{\hspace{1mm}}l@{\hspace{3mm}}c@{\hspace{1mm}}c@{\hspace{3mm}}c@{\hspace{1mm}}c@{\hspace{3mm}}c@{\hspace{1mm}}c@{\hspace{5mm}}c@{\hspace{2mm}}c@{\hspace{2mm}}c@{\hspace{5mm}}r@{\hspace{5mm}}c@{\hspace{3mm}}c@{\hspace{-2mm}}c}
\hline
\hline
Star\tablenotemark{a} &mode\tablenotemark{b} &\multicolumn{1}{c}{period} &$<B>$\tablenotemark{c} &($B$)\tablenotemark{c}&$<V>$&($V$) &$<I_{\mathrm C}>$& ($I_{\mathrm C}$)  &$A_B$\tablenotemark{d}   &$A_V$\tablenotemark{d}  &$A_{I_{\mathrm C}}$\tablenotemark{d}&$R$\tablenotemark{e}  &\multicolumn{3}{l}{\hspace{0mm}additional feature\tablenotemark{f}     } \\
&&\multicolumn{1}{c}{d}&mag&mag&mag&mag&mag&mag&mag&mag&mag&&&&\\
\hline
V13 &1O  &0.35072  &15.986 &15.953 &15.690 &15.673 &15.279 &15.271 &0.38-0.11 &0.29-0.09 &0.19-0.05&         &$f_{0.61}$&Bl&\\
     &F   &0.47950 &&&&&&                                          &0.97-0.40 &0.74-0.30 &0.50-0.20&        &&Bl& \\
     &2O  &0.28160 &&&&&&&                                          0.020     &0.019     &0.010    &        &&& PD \\
V44 &1O  &0.36812&  15.951 &15.928& 15.659& 15.646& 15.220& 15.215&0.25-0.14 &0.19-0.11&0.12-0.07 &         &&Bl&\\
     &F   &0.50377 &&&&&&&                                          0.87-0.34 &0.62-0.24 &0.40-0.19 &       &&Bl& \\
V68 &1O  &0.35597  &15.942 &15.909 &15.653 &15.635 &15.237 &15.230  &0.55      &0.40      &0.25     &        &$f_{0.61}$\tablenotemark{g}&&   \\
     &F   &0.47850 &&&&&&&                                          0.48      &0.35      &0.22      & 0.87    &&&   \\
V87 &1O  &0.35749  & 15.871 &15.853 &15.583 &15.573 &15.168 &15.164&0.49      &0.37      &0.23      &        &$f_{0.61}$&&\\
     &F   &0.48017 &&&&&&&                                          0.21      &0.16      &0.11      & 0.4    &&&  \\
V99 &1O  &0.3611   &15.940 &15.924 &15.653 &15.644 &15.223 &15.220 &0.52-0.13 &0.39-0.10 &0.26-0.06 &        &&Bl& \\
     &F   &0.4821  &&&&&&&                                          0.44-0.20 &0.38-0.15 &0.24-0.10  &         &&Bl& \\
V125 &1O  &0.34982 &15.939& 15.922& 15.665& 15.656& 15.262& 15.258 &0.54      &0.42      &0.26      & 0.02    &$f_{0.61}$&&\\
     &F   &0.4709  &&&&&&&                                         0.012      &0.012     &0.005     &        && \\                   
V166 &1O  &0.35672 &16.034 &16.007 &15.753 &15.736 &15.333 &15.326 &0.29-0.17 &0.23-0.14 &0.14-0.08 &        &&Bl&\\
     &F   &0.48504 &&&&&&&                                           0.99-0.36 &0.74-0.28 &0.47-0.18 &       &&Bl&PD\\
V200  &1O  &0.36102 &&&&&&&&&&&&&\\ 
     &F   &0.48530 &&&&&&&&&&                                                                         {\it 1.1}    &&&\\
V251$^*$  &1O  &0.3542 &&&&&&&&&&&&&\\
     &F   &0.4762 &&&&&&&&&&                                                                         {\it 0.9}    &&&\\
V252$^*$  &1O  &0.3361 &&&&&&&&&&&&&\\                                          
     &F   &0.4530 &&&&&&&&&&                                                                         {\it 0.95}   &&&\\ 
\hline
\end{tabular}
\tablenotetext{1}{ Asterisk denotes stars with photometry contaminated by signals of nearby variables.}
\tablenotetext{2}{ F: fundamental  mode; 1O:1st overtone; 2O:2nd overtone.}
\tablenotetext{3}{$<>$ and () denote magnitude and intensity averaged mean magnitudes, respectively.}\tablenotetext{4}{ Peak to peak amplitudes of data prewhitened for the other mode and the linear-combination terms. Maximum and minimum amplitudes are given for Blazhko stars.} 
\tablenotetext{5}{ Amplitude ratio relative to the amplitude of the 1O mode in $B$ band. Italics are for the non-calibrated data, transformed to magnitude with an uncertain zero point.} 
\tablenotetext{6}{ $f_{0.61}$: frequency at $f_{\mathrm{1O}}/f_{0.61}\approx 0.61-0.62$; Bl: Blazhko effect; PD:period doubling.} 
\tablenotetext{7}{ the $f_{0.61}$ frequency is detected only in the $V$ data.}  
\end{center}
\end{table*}

\begin{figure*}
\begin{center}
\includegraphics[width=18.5 cm]{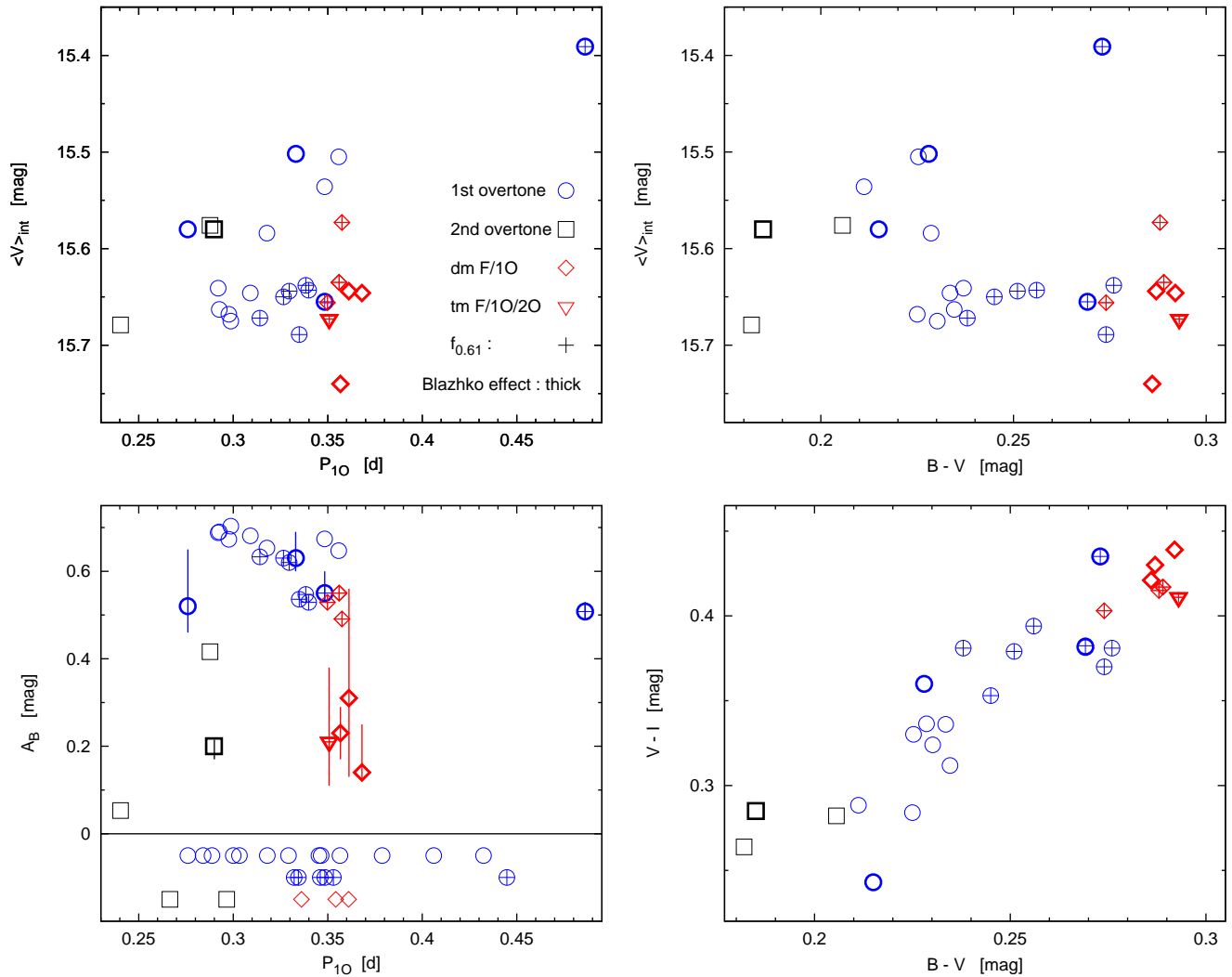}
\caption{Intensity-averaged mean $V$ magnitudes and peak-to-peak $B$ amplitudes vs period of overtone variables are plotted on the left-side panels. 1O, 2O variables, F/1O stars and the F/1O/2O  triple-mode variable,  V13, are denoted by different symbols. The appearance of the Blazhko effect and the additional frequency component at $f_{\mathrm {1O}}/f_{0.61}\approx 0.61-0.62$ ratio  is also indicated. The period distribution of stars with uncertain flux-magnitude transformation is shown in the bottom part of the $A(B)-P_{\mathrm{1O}}$ panel. For Blazhko stars the full range of the amplitude variation is indicated by a vertical line. The right-hand panels show intensity-averaged mean $V$ magnitude (top) and $V-I$  color (bottom) vs $B-V$.\label{p-va} }
\end{center}
\end{figure*}

\begin{figure*}
\begin{center}
\includegraphics[width=17 cm]{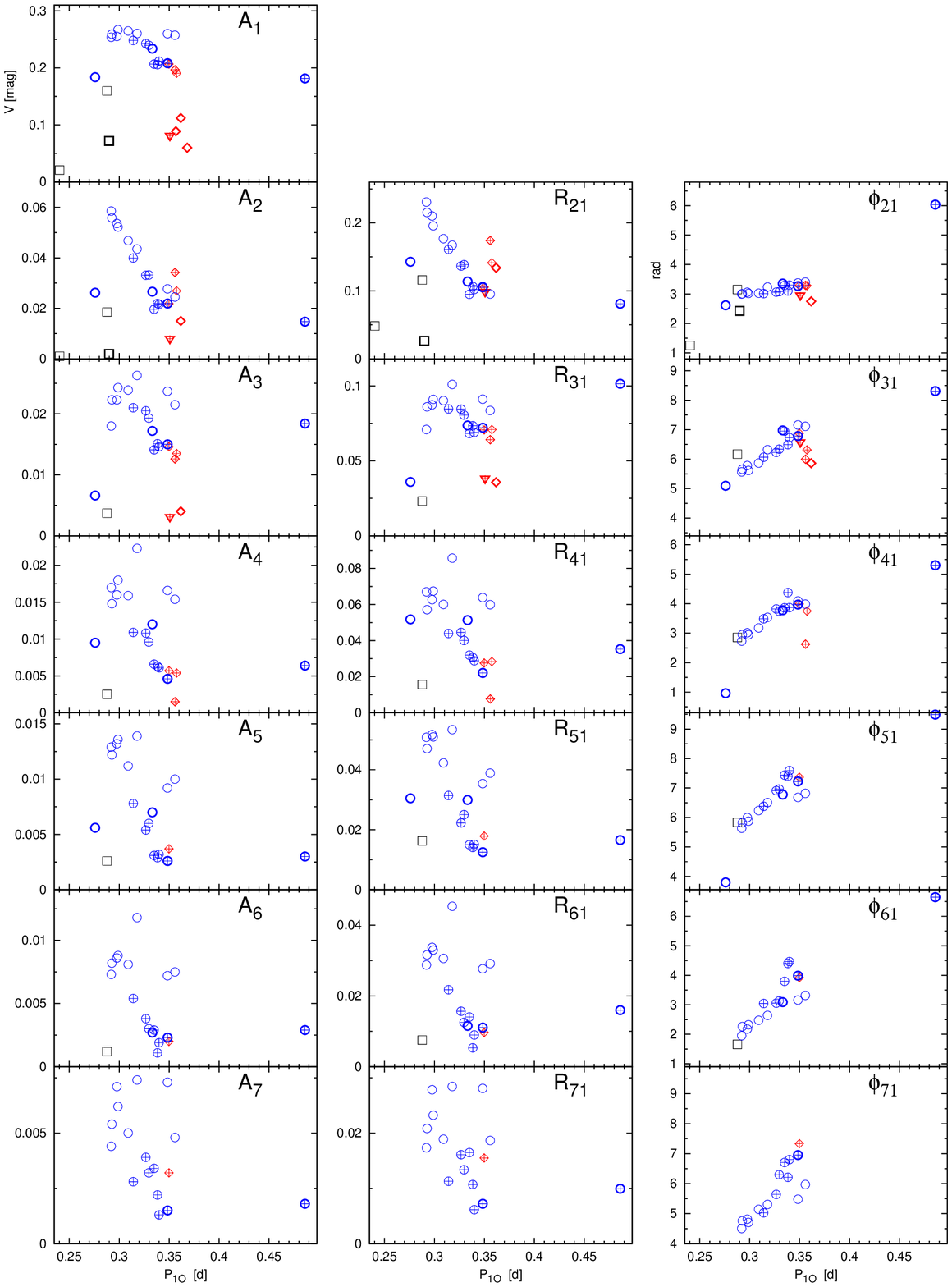}
\caption{Fourier amplitudes ($A_k$), amplitude ratios ($R_{k1}$) and phase differences ($\varphi_{k1}$) of the $V$ light curves of overtone  variables and the overtone mode of double-mode stars are shown. The symbols are the same as in Fig.~\ref{p-va}. Low-amplitude variables (second overtone, double-mode stars) do not appear in each plot, because they have detectable amplitudes only in the lowest harmonic orders.\label{four}}
\end{center}
\end{figure*}


\section{The photometric data: observations, data processing and analysis}\label{datasec}
An extended photometric campaign  was organized to observe M3 with the Konkoly Observatory 90/60~cm Schmidt telescope (Piszk\'estet\H o) equipped with a 4096x4096 Apogee CCD camera and Bessel $BVI_{\mathrm C}$ filters. About 1000 measurements were obtained in each band between January and July in 2012. The standard calibration processes of the CCD frames were performed using IRAF/CCDRED\footnote{{\sc IRAF} is distributed by the National Optical Astronomy Observatories, which are operated by the Association of Universities for Research in Astronomy, Inc., under cooperative agreement with the National Science Foundation.} tasks. The data were not corrected for the nonlinearity (less than 2\%) of the CCD system. 
As a consequence of the nonlinearity,  linear combinations of independent frequencies  may appear in the data as artifacts with mmag amplitude, and the measured amplitudes may be uncertain up to $1-2$\%. 
Both aperture (IRAF/DAOPHOT/PHOT) and Image Subtraction Method \citep[ISIS,][]{isis} photometry of the variables were performed. The light curves were transformed to the standard $BVI_{\mathrm C}$ system  using 23 bright, photometrically stable standard stars, covering a wide range in color, from the list of \citet{st00}. 

Time-series photometry of 52 overtone and double-mode RR Lyrae was obtained.
The flux-magnitude transformation of the results of the ISIS photometry needs accurate magnitudes on the reference frames. The PSF photometry (using  IRAF/DAOPHOT/ALLSTAR) on the reference frames was performed very circumspectly;  the neighboring stars of each variable were checked on SDSS images\footnote{http://www.sdss.org/.}. The $BVI_{\mathrm C}$ calibration of the light curves were accurate for 28 stars (Fig.~\ref{lcv}), but it remained uncertain because of crowding for 24 variables.
Besides of the pulsation period, additional properties of the light curves were detected in some of these non-calibrated flux time series. Therefore, the  magnitude transformation of the $B$-filter flux curves of these 24 stars was also performed (Fig.~\ref{lcb}) with reference magnitude values that yielded  mean magnitudes matching the horizontal-branch magnitude of the cluster, just for the visualization of the light curves. The time-series photometry obtained using the image subtraction method is sensitive to the contamination of only nearby variables. The frequencies of close variables in the  neighborhood appeared with low amplitude in the spectrum of some of the variables with non-calibrated light curves, but these signals were removed from the data.
No flux curve could be derived for five overtone variables (V152, V155, V178, V266, and V269) because their  ISIS photometry  gave unreliable results due to blending with another variable.
The CCD time series of the variables are available from the electronic edition of this article; Table~\ref{photmag} and Table~\ref{photflux}  give samples of the magnitude calibrated $B$, $V$, ${I_{\mathrm C}}$, and the $B$-filter flux data regarding  their form and content. These data are referred as D12 in the paper.

The light curves were analyzed using  the program packages MUFRAN \citep{mufran} and LCfit \citep{nl}. Independent frequencies were selected in successive prewhitening steps if their amplitudes were larger than 4$\sigma$ at least in 2 bands. For linear-combination terms a less strict criterion was used. 

The long-term behavior and the changes of some special properties of the stars were checked using the CCD data published in K98, C01, H05, B06 and J12  and in the photographic data collected in J12. All the previous photometric data are magnitude homogenized and merged.

\section{The light curves, variable types} \label{lcsec}
Table~\ref{rrc.dat} and Table~\ref{rrd.dat} summarize the main characteristics of the light curves of the overtone and the double- (multi-)mode variables,  respectively, based on the D12 data. Pulsation mode, period, magnitude-, and intensity-averaged $B, V, I_{\mathrm C}$ brightnesses, amplitudes, and remarks on any additional property are listed in the columns. The amplitude-ratio of the radial modes in $B$ band is also given for non-Blazhko RRd stars in Table~\ref{rrd.dat}. The stars for  which the photometry is contaminated by signals from nearby (within 6") variable stars are denoted by asterisk in Tables~\ref{rrc.dat} and \ref{rrd.dat}. 

The panels in Fig.~\ref{p-va} show the location of the stars on the $V$-period, $A(B)$-period, $V$-$(B-V)$ and $(V-I)$-$(B-V)$ planes. The double-mode (red triangle and diamond), first overtone (blue circle) and second-overtone (black square)  stars are located in separated groups with some overlapping in each plot. 
The extra properties of the stars, the  Blazhko effect and the $f_{0.61}$ additional frequency are also indicated in the figure.

The Fourier amplitudes, amplitude ratios and phase differences of the $V$ light curves are shown in Fig.~\ref{four}. The  same symbols  are used for distinguishing stars with different properties as in Fig.~\ref{p-va}. The location of different-type variables in Figs.~\ref{p-va} and ~\ref{four} are discussed in the next corresponding  sections.

\subsection{Multi-mode pulsations, additional frequencies}\label{mmasec}
The light curves of more than half of the studied stars are not stable; their Fourier spectra show extra frequency components. For example, an additional frequency component appears in 18 stars at the frequency ratio  $f_{{\mathrm{1O}}}/f_{0.61} \approx 0.61-0.62$ (RR$_{0.61}$ stars).
Period doubling (PD) --  manifested as signal at half-integer frequency ratio to the original one -- is detected in two cases (V13 and V166). When the additional frequency appears close to the overtone frequency (and its harmonics), the Blazhko effect (Bl) is supposed to be observed, even if only one close frequency component is detected (V126, V166, and V203). Though, alternatively, the additional signal may correspond to a non-radial mode in these latter cases. 

Ten double-mode variables are in the sample, one of them is, in fact, a triple-mode star.  Besides the fundamental (F) and the first-overtone (1O) modes, the second-overtone mode (2O)  [the identification is based on  the $P/P_{\mathrm F}=0.587$ period ratio]  also appears in the spectrum of V13. If the 2O identification of the third mode is correct,  V13 is the first RR Lyrae star showing pulsation with the first three radial modes simultaneously.

The occurrence of the different types of phenomena are frequently interlaced. The most complex variation is observed in the triple-mode star, V13, which shows Blazhko effect of the F and 1O radial modes, period doubling of the 2O mode and the $f_{0.61}$ component as well. About half of the RRd stars show the Blazhko effect and their $P_{\mathrm{1O}}/P_{\mathrm F}$ period ratios  are anomalous.

 In contrast with the fast and irregular period changes often detected in RRc stars, the period changes of overtone variables in M3 are marginal or modest. Irregular, rapid period changes were detected only of two RRc stars (V12 and V70) and of some RRd stars  in the period-change study of the RR Lyrae stars of M3 (J12). Therefore, the possibility that the detected additional features were the artifacts of period changes in a 200-d long time-series photometry can be excluded in most of the cases. The possible role of the period changes of V70 is discussed in Section~\ref{blnotesec}, and the period changes of double-mode stars are summarized in Section~\ref{dmsec}.

\subsection{Evolved and second-overtone variables}\label{evolsec}
The overtone variables (V70, V170, and V261) with the longest periods are close to the short-end of the period range of the fundamental-mode variables ($P=0.40-0.45$~d) and have light curves showing anomalous shape. Their rising branch is longer than of typical ones and the bump appears at the middle of it, while that generally precedes maximum light in normal RRc stars slightly. Therefore, the RRc status of these highly evolved variables might be questionable. However, two of these stars (V70 and V261) show a secondary periodicity with 0.61 period ratio. The fact, that this type of multi-periodicity appears in many other RRc stars in M3, and exclusively in first-overtone variables according to other studies (see details in Section~\ref{f6sec}) supports the RRc status of these stars.

The identification of three 2O variables (V105, V203, and V290) is based on their short period, small-amplitude, smooth light curves and extreme color indices. 
Though its period and amplitude may indicate 2O status of V168 ($P=0.276$ d) too, as this star shows a strong bump preceding maximum light in the large-amplitude phase of its modulation, which feature is typical in RRc stars, thus its 2O classification is rejected. 
Second-overtone status of two more stars with no color information  (V188 and V292) is also proposed based on their  short periods and small-amplitude smooth light curves.

\section{Additional frequencies at 0.61 frequency ratio}\label{f6sec}

\begin{table}
\begin{center}
\caption{Parameters of the $f_{0.61}$  frequency components.\label{af.dat}}
\begin{tabular}{@{\hspace{0mm}}l@{\hspace{2mm}}l@{\hspace{3mm}}l@{\hspace{2mm}}c@{\hspace{1mm}}c@{\hspace{2mm}}c@{\hspace{0mm}}c}
\hline
\hline
Star &  $f_{\mathrm{1O}}$ &  $f_{\mathrm 0.61}$ &$f_{\mathrm{1O}}/f_{\mathrm 0.61}$ &${A_{0.61}}$\tablenotemark{a} &${A_{0.61}/A_{{\mathrm{1O}}}}$\tablenotemark{b} &\multicolumn{1}{c}{Data}\tablenotemark{c}\\
&d$^{-1}$&d$^{-1}$&& mag&&\\
\hline
V13       &2.8513  & 4.6457 & 0.6137 &0.006 &0.075 & D12\\
V37       &3.0615  & 4.977  & 0.6151 &0.006 &0.025 & D12\\
          &        & 4.965  & 0.6166 &0.004 &0.018 & D12\\
          &3.0615  & 4.9660 & 0.6165 &0.007 &0.028 & H05+B06   \\
V56       &3.0340  & 4.944  & 0.6137 &0.009 &0.036 & D12\\
          &        & 4.911  & 0.6137 &0.005 &0.019 & D12\\
          &3.0341  & 4.9475 & 0.6133 &0.007 &0.028 & J12       \\
          &3.0339  & 4.9323 & 0.6151 &0.009 &0.035 & H05+B06   \\
V68       &2.8092  & 4.5717 & 0.6145 &0.005 &0.028 & D12\\
V70       &2.0573  & 3.3115:& 0.6213 &0.004 &0.024 & D12\\
V75       &3.1839  & 5.1595 & 0.6171 &0.006 &0.023 & D12\\
          &3.1840  & 5.1506 & 0.6182 &0.008 &0.032 & J12       \\
V87       &2.7973  & 4.5288 & 0.6177 &0.006 &0.034 & D12\\
V97       &2.9857  & 4.879  & 0.6147 &0.005 &0.026 & D12\\
          &        & 4.868  & 0.6133 &0.005 &0.025 & D12\\
          &        & 4.857  & 0.6119 &0.005 &0.024 & D12\\
          &2.9856  & 4.8597 & 0.6144 &0.010 &0.050 & D12a \\
          &2.9857  & 4.8703 & 0.6130 &0.005 &0.022 & D12b\\
          &2.9857  & 4.8370 & 0.6173 &0.008 &0.041 & H05+B06   \\
V125      &2.8586  & 4.654  & 0.6142 &0.006 &0.028 & D12\\ 
          &        & 4.640  & 0.6161 &0.005 &0.022 & D12\\ 
          &        & 4.590  & 0.6229 &0.005 &0.025 & D12\\
          &2.8586  & 4.6343 & 0.6168 &0.006 &0.031 & J12       \\
          &2.8587  & 4.6344 & 0.6168 &0.009 &0.047 & H05+B06   \\      
V126      &2.8702  & 4.604: & 0.6234 &0.005 &0.023 & D12\\
V132      &2.9424  & 4.7913 & 0.6141 &0.011 &0.053 & D12\\
          &2.9425  & 4.7497 & 0.6195 &0.011 &0.057 & B06       \\
V147      &2.8863  & 4.6947 & 0.6148 &&{\it{0.028}}& D12\\
V208      &2.9552  & 4.8058 & 0.6149 &0.013 &0.065 & D12\\
V209      &2.8712  & 4.6800 & 0.6135 &&{\it{0.033}}& D12\\
V210      &2.8333  & 4.6185 & 0.6135 &&{\it{0.044}}& D12\\
V253      &3.0066  & 4.8720 & 0.6171 &&{\it{0.079}}& D12\\
V259      &2.9984  & 4.8773 & 0.6148 &&{\it{0.026}}& D12\\
V261      &2.2485  & 3.6237 & 0.6205 &&{\it{0.069}}& D12\\
\hline
\end{tabular}
\tablenotetext{1}{Fourier amplitude in $V$ band.}
\tablenotetext{2}{Amplitude ratio in $V$ band. Italics denote $B$-band amplitude ratio of non-calibrated data.}
\tablenotetext{3}{D12a and D12b denotes first and second parts of the D12 data.}
\end{center}
\end{table} 

\begin{figure}
\includegraphics[width=9 cm]{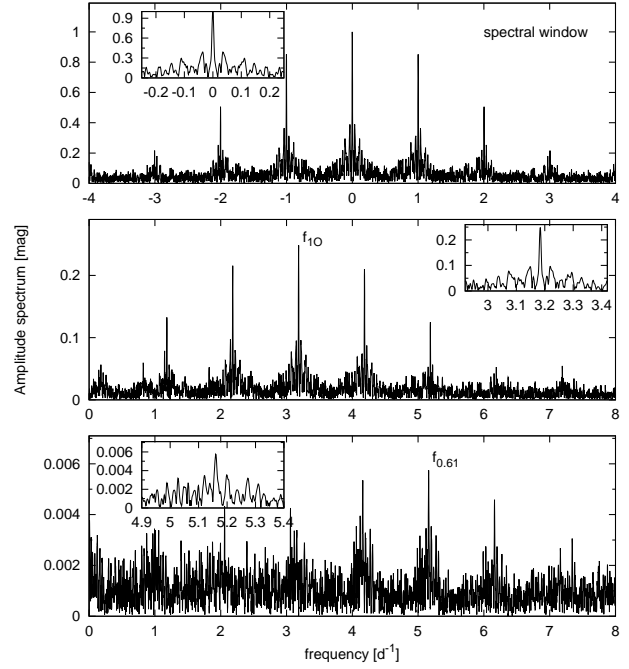}
\caption{Spectral window (top), amplitude spectrum, and spectrum of the data prewhitened for the overtone-mode frequencies of V75 in the $V$ band. The insets enlarge the main peaks. The detection of the $f_{0.61}$ frequency in the prewitened spectrum of V13 is shown in Section~\ref{pdsec}.\label{75sp}}
\end{figure}

\begin{figure}
\includegraphics[width=9 cm]{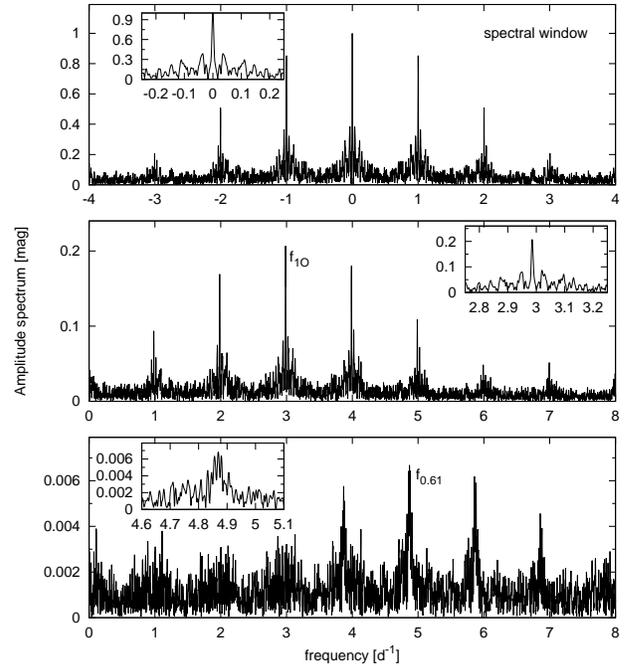}
\caption{The same spectra as in Fig.~\ref{75sp} for V97. The $f_{0.61}$ frequency is a complex multiplet in this case.\label{97sp} }
\end{figure}

The  $f_{0.61}$ frequency ($f_{0.61} \approx f_{\mathrm{1O}}/0.61$) is detected in 14 RRc and in four double-mode stars in M3. Thus, the phenomenon is observed in 38 and 40\% of RRc and RRd stars, respectively. Fig.~\ref{75sp} and Fig.~\ref{97sp} show the spectral windows, the amplitude spectra and the prewhitened spectra of V75 and V97, as examples for the detection of the $f_{0.61}$ frequency in the D12 data.  

Table~\ref{af.dat} lists the RR$_{0.61}$ stars and gives their frequencies, frequency ratios,  $A_{0.61}$ amplitudes in $V$ band and the $A_{0.61}/A_{\mathrm {1O}}$ amplitude ratios. The previous CCD data (C01, K98, H05, B06, and J12) of these variables have also been re-investigated and the $f_{0.61}$ frequency was detected in six stars in the archive photometric data; these results are also given in Table~\ref{af.dat}.

\begin{figure*}
\includegraphics[width=18.5 cm]{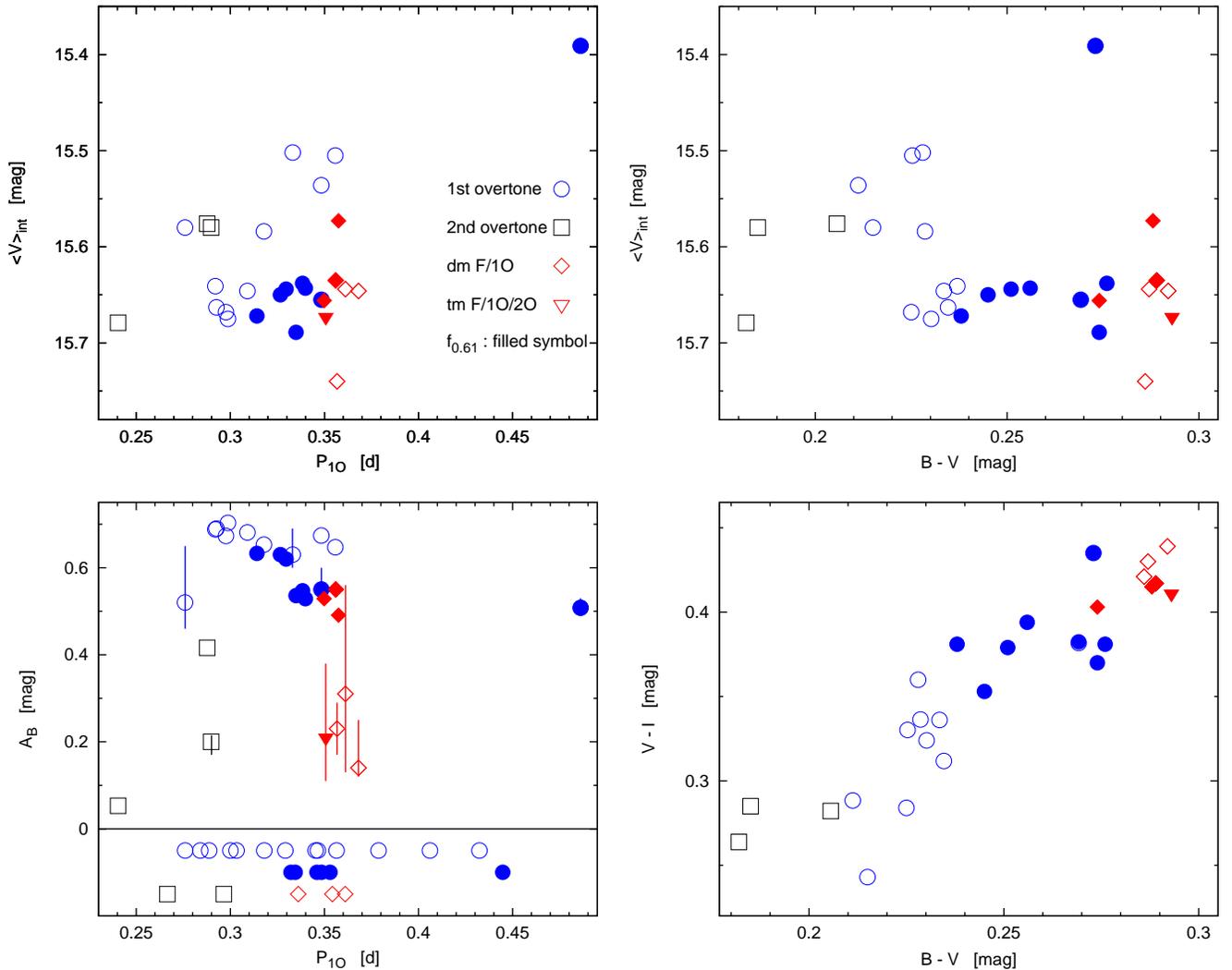}
\caption{The same as in Fig.~\ref{p-va}. RR$_{0.61}$ stars are shown by filled symbols.\label{p-va.f6} }
\end{figure*}

\subsection{Stability of the $f_{0.61}$ frequency}\label{f6stabsec}

The $f_{0.61}$ frequencies detected in the archive CCD  and  D12 data  differ from each other in most of the cases and the differences are too large to be explained by the $\sim$$0.0001-0.001$~d$^{-1}$ uncertainty of the frequencies.

The $f_{0.61}$ frequency is a discrete, stable signal in V75, but this component has a complex, multiple structure in V97, indicating probable amplitude and/or phase changes of the signal. A complex structure of the $f_{0.61}$ frequency is detected in the residual spectra of four variables in the D12 data. The stability of the frequency solution of these stars is checked on V97. The first and second parts of the observations are analyzed separately, and these results are also given in Table~\ref{af.dat}. The frequency and the amplitude of the $f_{0.61}$ component are significantly different in the two parts of the data. This, and the differences between the results of the D12 and the archive data imply that, instead of real multiplets, amplitude and/or frequency changes of the $f_{0.61}$ component should be behind its complex appearance. This is in contrast with the stability of the radial-mode frequency of RR$_{0.61}$ stars. While as large as $\sim$$1$\% differences are detected in the frequency values of the $f_{0.61}$ component, the variation in the frequency of the radial-mode is lower than 0.01\% (see Table~\ref{af.dat}).

\subsection{On the light curves of RR$_{0.61}$ stars}\label{f6lcsec}

The appearance of the $f_{0.61}$  frequency in different groups of stars are indicated in Figs.~\ref{lcv} and \ref{lcb}. The light curves of these variables are more sinusoidal than the light curves of overtone variables without any additional feature. While a pronounced bump characterizes the light variation of normal RRc stars just preceding light maximum, the bump in these stars is marginal if it exists at all.  The light curves of the 1O mode of double mode  stars have a similar shape as RR$_{0.61}$ stars have, irrespective of whether or not the $f_{0.61}$ frequency is detected in their spectra (see also Fig.~\ref{lcd} in Section~\ref{dmsec}). 

According to the best-quality light curves shown in Fig.~\ref{lcv},  there is no star contradicting this scheme.  Each RR$_{0.61}$ star displays a sinusoidal light curve with a strongly reduced bump, similar to the overtone-mode light curve of double-mode stars, while the light curves of normal RRc stars are asymmetric and their bump is pronounced. The only exception is V70, which is a highly evolved, over-luminous, long-period RRc star.

The non-calibrated light curves of variables affected by crowding shown in Fig.~\ref{lcb} follow the same pattern with some exceptions. Each RR$_{0.61}$ star of this sample fits to this scheme, their bumps are marginal. However,  similar, sinusoidal light curves are found among normal RRc stars, as well. Because of the larger noise level, the detection limit of the additional frequency is larger  in these cases than in the best-quality light curves. Therefore, it cannot be ruled out that these `outliers' may, in fact, exhibit the `0.61' feature as well, but the quality of the light curve does not allow its detection. There is an evolved, long-period star in this sample (V170), too, which shows the $f_{0.61}$ frequency.

\begin{figure}
\includegraphics[width=8.2 cm]{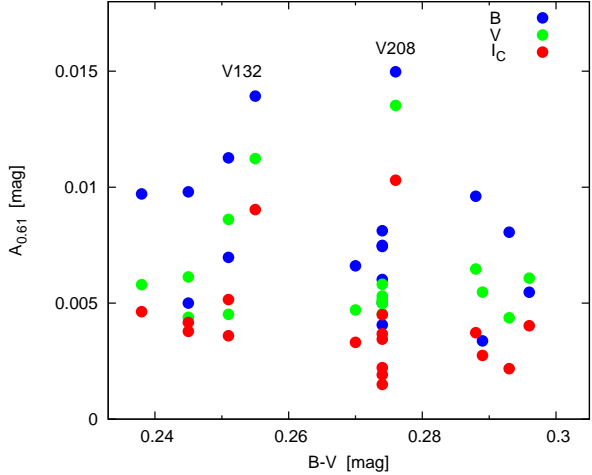}
\caption{$B$-, $V$-, $I_{\mathrm C}$-band  amplitudes  of the $f_{0.61}$ frequency component vs $B-V$. No amplitude decrease of the $f_{0.61}$ frequency towards smaller color indices is detected. The two largest $f_{0.61}$-amplitude stars are labeled. \label{abv}}
\end{figure}

\begin{figure}
\includegraphics[width=8.2 cm]{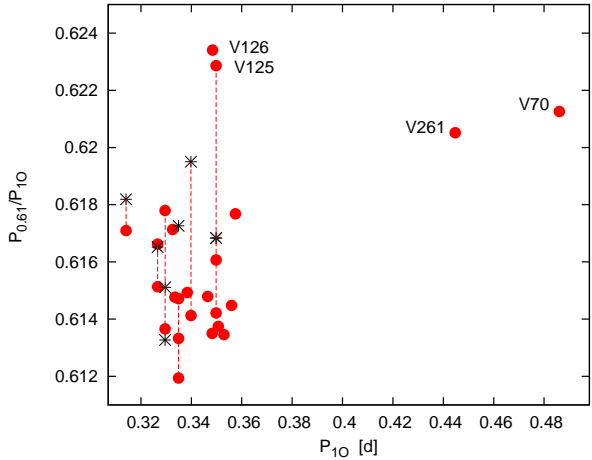}
\caption{Petersen diagram, period ratio vs $P_{\mathrm {1O}}$ of variables showing the $f_{0.61}$ frequency component. The detected frequencies of the same star are connected with dashed lines (see data in Table~\ref{af.dat}). Results for archive CCD data are shown by asterisk.\label{f6} }
\end{figure}

\begin{figure}
\includegraphics[width=8.2 cm]{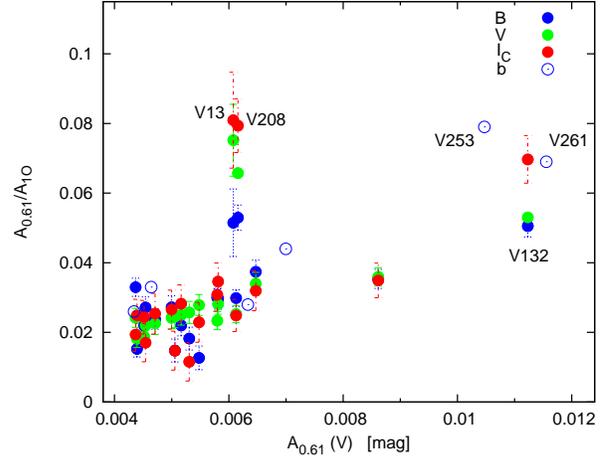}
\caption{ $B$-, $V$-, $I_{\mathrm C}$-band and  non-calibrated $B$-filter amplitude ratios ($A_{0.61}/A_{\mathrm{1O}}$) vs the Fourier  $V$ amplitude of the $f_{0.61}$ frequency  ($B$-filter amplitude of the non-calibrated data). For multiple  $f_{0.61}$ signals each component is shown.\label{a6}}
\end{figure}

\begin{figure*}
\includegraphics[width=17.5 cm]{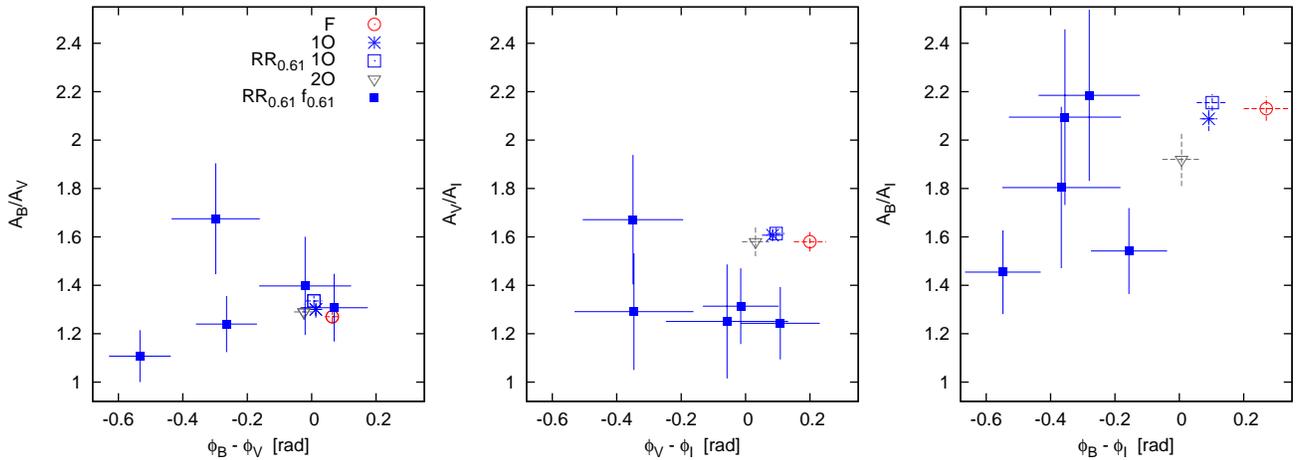}
\caption{The amplitude ratios and phase differences of single-mode F, 1O, and 2O variables, the first-overtone mode of RR$_{0.61}$ stars, and the $f_{0.61}$ frequency component in the $B$, $V$  and $I_{\mathrm C}$ bands are plotted. With the exception of the $f_{0.61}$ components, the mean values of the parameters and the standard deviations of the samples are shown. 
Because of their small amplitudes, the amplitude ratios and phase differences of the $f_{0.61}$ frequency components  have large errors, therefore, results only of the 5 largest S/N components are shown. The formal errors, which, however, may underestimate the true uncertainties of the $f_{0.61}$ components, are indicated. \label{6}}
\end{figure*}

The Fourier parameters of the light curves (Fig.~\ref{four}) also reflect the above described diversity. While mono-periodic RRc stars have detectable amplitudes  up to  ten or even larger harmonic orders in $V$ band,  we can find harmonics only up to $6-8$ with amplitudes  above the noise limit in stars showing the $f_{0.61}$ frequency. The Fourier amplitudes and amplitude ratios of the RR$_{0.61}$ stars are systematically smaller than that ones of normal RRc stars. Only  double mode, Blazhko, and 2O stars have amplitudes and amplitude ratios as low as RR$_{0.61}$ stars have over the third harmonic order. The phase differences of the  RR$_{0.61}$ stars do not show any peculiar behavior, they follow the trend defined by other RRc stars.

\subsection{Parameters of RR$_{0.61}$ stars}\label{f6parsec}
The location of RR$_{0.61}$ stars   is also special in the panels shown in Fig.~\ref{p-va.f6}. These plots are the same as in Fig.~\ref{p-va}, but the RR$_{0.61}$ stars are marked by filled   symbols for clarity. The  $f_{0.61}$ frequency component is not detected in any of the stars with period shorter than 0.31 d, while it appears in the residual spectrum of each, not over-luminous RRc star at longer periods, and in four double-mode stars. None of normal-period, bright, evolved RRc stars shows the  $f_{0.61}$ frequency, but it is detected in two of the five, highly evolved stars. The periods of these 1O pulsators (V70, $P=0.486$~d and V261, $P=0.445$~d) are longer than the 1O period of double-mode stars; they are in the period range of the shortest-period fundamental-mode variables. Although the magnitude calibration of V261 is uncertain, as its period and light-curve shape is similar to those of V70, the brightest, longest-period RRc star in M3,  it is very probable that there are no significant differences between the luminosities and temperatures of V261 and V70.

The separation of RRc and  RR$_{0.61}$ stars is the most pronounced in the $B-V$ - $V-I$ panel of Fig.~\ref{p-va.f6}. It seems that the $f_{0.61}$ phenomenon occurs only in  the hotter half of the temperature  range of overtone and double-mode stars in M3. There are only three double-mode stars (V44, V99, and V166) in the hotter half of the complete RRc/RRd sample  where  the $f_{0.61}$ frequency is not detected. However, these stars exhibit complex Blazhko modulation of both modes and the mean amplitude of their 1O mode is only about the half of the amplitude of the other RRc stars and of RRd stars with  dominant overtone mode. Supposing that the $A_{\mathrm {1O}}/A_{0.61}$ amplitude ratio  is not extremely large (as in the case of V13, the only RRd star with  dominant fundamental mode showing the $f_{0.61}$ component), the estimated amplitude of the $f_{0.61}$ frequency might be below the detection limit in these three RRd stars. The $f_{0.61}$ frequency is not detected in the RRd stars with uncalibrated, noisy light curves either, but the $f_{0.61}$ frequency may be hidden in the noise in these stars, too. Therefore, we cannot exclude the possibility that the  $f_{0.61}$ frequency is a common property  of RRd  stars and of the hotter half of the RRc sample. In contrast, none of the RRc stars with $B-V\lesssim0.24$ and $V-I\lesssim0.35$ mag shows the $f_{0.61}$ signal.

Fig.~\ref{abv} shows the amplitudes of the $f_{0.61}$ frequency components plotted against the $B-V$ color index. No systematic amplitude decrease of the $f_{0.61}$ frequency  is detected with decreasing color index. Consequently, the lack of RR$_{0.61}$ stars at hotter temperatures as shown in  the $B-V$ vs $V-I$ panel of Fig.~\ref{p-va.f6} cannot be explained by a systematic reduction of the amplitude of the $f_{0.61}$ component towards hotter temperatures.

Fig.~\ref{f6} plots the observed  $P_{0.61}/P_{\mathrm {1O}}$ period ratios of the 18 RR$_{0.61}$  stars. Each  component  of the stars with multiple $f_{0.61}$ frequencies and results for the archive CCD data are shown. The period ratios fall in a very narrow range between 0.612 and 0.623. The four largest values are detected in peculiar stars. Two of them, V70 and V261 are  evolved, extremely long-period variables. The additional frequency is a strongly varying signal with three $f_{0.61}$ components appearing in the whole data set of V125. The period ratios of two of these components and those obtained from the H05, B06 and J12 data fall in the $0.614-0.617$ range. The largest period ratio is detected in V126. The $f_{0.61}$ component is one of the weakest in this star, thus the frequency solution is somewhat uncertain in this case.

 Fig.~\ref{a6} shows the $A_{{0.61}}/A_{{\mathrm {1O}}}$ amplitude ratios  in $B$, $V$,  and $I_{\mathrm C}$ bands (and for the non-calibrated $B$-filter data) derived from the D12 data. Most of the amplitude ratios are in the $0.02-0.04$ range. The amplitude ratios of five stars (V13, V132, V208, V253, and V261) are even larger, they are about $0.05-0.08$.  The increased  $A_{{0.61}}/A_{{\mathrm {1O}}}$ amplitude ratios are the consequences of either the relatively large amplitude of the  $f_{0.61}$ component or the reduced amplitude of the overtone mode as in the case of V13. No general, systematic difference between the  amplitude ratios in the $B$, $V$, $I_{\mathrm C}$ bands is seen for most of 
the $f_{0.61}$ components. However, the three large amplitude-ratio stars with $BVI$ photometric data (V13, V132, and V208) show an anomalous tendency, their  $A_{{0.61}}/A_{{\mathrm {1O}}}$ amplitude ratio is the largest in the $I_{\mathrm C}$ and the smallest in the $B$ band. However, taking into account  that the formal errors may  underestimate the true uncertainties seriously, this result has to be taken with caution. 

The amplitude ratios and phase differences of single-mode F, 1O, 2O variables, and the first-overtone mode of RR$_{0.61}$ stars (mean values and the standard deviations of the samples), and of the $f_{0.61}$ frequency component  are plotted in Fig.~\ref{6} for the $B$, $V$,  and $I_{\mathrm C}$ bands. The amplitude ratios and phase differences of the $f_{0.61}$ components  have large errors  because of their small amplitudes.  Therefore, results are shown only for the five best, smallest-error components.
There is no detectable difference between the positions of the overtone mode of the mono-periodic and the  RR$_{0.61}$ stars shown  by an asterisk and a square, respectively. The phase differences of the F and 2O modes are at slightly larger and smaller values than the phase differences of the 1O. The amplitude ratios and phase differences  of  the best S/N-ratio  $f_{0.61}$ components form a separate group in the $V$,$I$ and $B$,$I$ planes  with no overlap with the regions of the radial modes. Although the errors even of the best signals are significant, the homogeneous location of these components implies that the amplitude ratios and phase differences of the $f_{0.61}$ frequency and of the radial modes may indeed be different.
However, more accurate multicolor observations of  RR$_{0.61}$ stars are needed to locate the positions of the  $f_{0.61}$ component in the amplitude-ratio and phase-difference planes undoubtedly.


\subsection{Comparison with other results}\label{f6compsec}
The  $f_{0.61}$ frequency  has been detected in several overtone RR Lyrae stars previously. A complete list of these objects was given  by \cite{mo15} recently. Moreover, \cite{netzel} identified 147 Galactic bulge objects in the OGLE-III photometry which all showed the peculiar frequency component in the $0.60-0.65$ period-ratio range. The period ratios of the OGLE sample defined two sequences, one at 0.613 and a less populous group at 0.631 ratio. \cite{mo15} argued that this frequency cannot be identified with any radial mode. They have concluded, that most probably the $f_{0.61}$ frequency corresponds to a strongly trapped, unstable, high-degree ($l\geq6$) non-radial mode. 

Color information may help  mode identification of non-radial modes \citep[see e.g.][]{g04}, however, RR Lyrae stars still lack appropriate modeling to exploit this possibility.

Similar period ratios were detected in numerous Magellanic Cloud 1O Cepheids \citep{so08,mk08,mk09,so10}. The period ratios of these stars fell on three parallel lines, separated by about 0.02 in period ratios in the $P/P_{\mathrm{1O}} - {\mathrm{log}}P_{\mathrm{1O}}$ Petersen diagram.  In a theoretical work, \cite{dz12} concluded that in Cepheids `the only unstable modes that may reproduce the observed period ratio are $f$-modes of high angular degrees ($l=42-50$)'. This hypothesis predicts very large ($35-70$~kms$^{-1}$) broadening of the spectral lines of these stars, which has not been verified observationally yet.

RR Lyrae stars are known to be slow rotators \citep{peterson,pc} with microturbulent velocity of about $2-3$~kms$^{-1}$ \citep[][and references therein]{ggp} and no significant line broadening of any RR Lyrae stars has been detected previously.

Spectroscopic data of M3 variables were also collected  by  Hectochelle@MMT \citep{mmt} with $\sim$20000 spectral resolution in the $5150-5300$~\AA   wavelength range parallel with the photometric observations in 2012. The details of the observations and data reduction process will be published in a separate paper.

\begin{figure}
\includegraphics[width=8.9 cm]{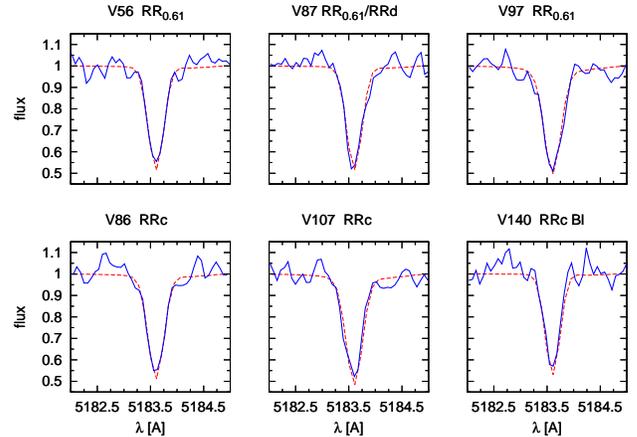}
\caption{The Mgb (5183.604 \AA) absorption line in the spectra of 6 RRc/RRd stars at the pulsation phase around minimum brightness. Stars showing the $f_{0.61}$ frequency are in the top panels. The line profiles are fitted with synthetic spectra of appropriate physical parameters and of rotational velocities of 5 kms$^{-1}$ ($v$sin$i$) for each star.  No systematic difference is detected between the broadenings of the line profiles of RR$_{0.61}$  and other 1O stars.\label{lw}}
\end{figure}

Six overtone RRL stars belonged to the targets of the spectroscopic observations. Three of them, the RRc stars V56, V87 and the RRd star, V97 show the $f_{0.61}$ frequency. The others are normal RRc stars (V86 and V107),  and one of them, V140 shows strong Blazhko modulation. We have checked the line broadenings of these stars in pulsation phase near to minimum brightness. Fig~\ref{lw} shows the line profiles of the Mgb (5183.604 \AA) absorption line, the deepest, non-blended  line in the spectra of the 6 RRc/RRd stars and the best fitting models using the alpha enhanced synthetic spectra ($Z=-1.5$) of the Asiago stellar library \citep{munari}. The atmospheric parameters ($T_{\mathrm {eff}}$, log$g$) were estimated from the observed colors assuming $E(B-V)=0.01$~mag interstellar reddening using the synthetic colors published in \cite{kurucz}. Uncertainties of the physical properties account for about $1-2$~kms$^{-1}$ errors of the results.

The broadenings of the line profiles are estimated to be about 5~kms$^{-1}$ for all the six stars (Fig.~\ref{lw}). The line broadening of the three RR$_{0.61}$ stars are thus the same as for RRc stars not showing the $f_{0.61}$ component. The lack of any  systematic difference  between the broadenings of the line profiles of RR$_{0.61}$  and other 1O RRL stars in M3 contradicts the high-order $f$-mode explanation of the $f_{0.61}$ frequency proposed by \cite{dz12}.


Amplitude and phase variations of the $f_{0.61}$ component  on a time scale of $10-200$~d was detected in the Kepler data \citep{mo15}. However, the Kepler data revealed very low-amplitude (tenths of mmag) variations of the radial mode on similar timescale, as well. Such variations are below the detection limit of the D12 data.

The light curves of the Kepler RRc stars with the $f_{0.61}$ frequency component are also sinusoidal or show only a marginal bump  \citep[see fig.~2 in][]{mo15}, as in M3. Although no H emission in the pre-maximum phase of RRc stars has been reported, the pre-maximum bump in RRab stars is related to a strong  atmospheric shock \citep[][and references therein]{gf}. The lack of a pronounced bump on the light curves of RR$_{0.61}$ stars might indicate that the formation of atmospheric shocks and the excitation of the $f_{0.61}$ frequency are  mutually exclusive.

The amplitude ratio of the  $f_{0.61}$ frequency relative to the amplitude of the radial mode is in the  $0.02-0.08$ range in M3. The amplitude ratios are in a similar range in all the other samples of stars showing the $f_{0.61}$ frequency. They are $0.01-0.04$, $0.02-0.05$, and $0.02-0.08$ in Cepheids \citep{so10}, in the RRc stars observed by the Kepler space telescope \citep{mo15}, and in the Galactic bulge \citep{netzel}, respectively.  The $f_{0.61}$ frequency is thus a relatively small-amplitude signal according to each of these studies.

Each of the four RRc stars observed by the Kepler mission shows the $f_{0.61}$ frequency, moreover,  period doubling of this component are detected in these stars \citep{mo15}.  Half-integer frequency of the $f_{0.61}$ component appears in four of the 147 Galactic bulge RR$_{0.61}$ stars, too \citep{netzel}. The amplitudes of these half-integer frequencies are smaller than 0.003 mag with one exception, the amplitude of the $f_{0.61}/2$ frequency of KIC 5520878 is 0.006 mag.
No $kf_{0.61}/2$ half-integer frequency is detected in M3, however, signals with amplitude smaller than $~0.004$~mag are below the detection limit of the D12 data.

The $f_{0.61}$ frequency is detected in $38$\% of the RRc/RRd stars in M3. \cite{mo15} found that almost every RRc and RRd star observed from space showed the $f_{0.61}$ frequency, consequently the phenomenon must be very common. Concerning Cepheids, $\sim$$9$\% of the 1O pulsators showed the $f_{0.61}$ frequency in the SMC \citep{so10},  but none of the shortest period ones  (log$P_{\mathrm {1O}}<0.056$) as noticed by \cite{dz12}. It seems that in this sample there is a tendency that the phenomenon does not occur in the hottest stars.

About 20\% of the RR$_{0.61}$ stars are bona-fide double-mode pulsators in M3. 
There are 4 and 2  RRd stars in the samples of  23 field and globular cluster \citep{mo15}, and 147 Galactic bulge  RR$_{0.61}$ stars \citep{netzel}, respectively. Thus the frequency of double-mode pulsators in these data sets are 17, and 1.4\%. The very low rate of RRd stars showing  the $f_{0.61}$ frequency in the Galactic bulge is the consequence of the fact that the occurrence rate of RRd stars  among overtone variables in the Galactic bulge \citep{so11} is only $\sim$$2$\%.

The homogeneity of GC data allows  to determine the positions of RR$_{0.61}$ stars relative to other overtone variables using their measured parameters. Each RRc star close to the blue edge of the double-mode region -- based on their periods and colors -- and four RRd stars show the $f_{0.61}$ frequency. It is also detected in some highly evolved, bright, long-period variables, which temperature (indicated by their colors) fall in the temperature range covered by the other RR$_{0.61}$ stars. The similarity of the light-curve shapes of RR$_{0.61}$ and RRd stars (with reduced or missing bump) and their physical parameters (based on the colors and periods) indicates that the  $f_{0.61}$ phenomenon might be strongly connected to the double-mode state. Maybe, the complex dynamics of the observed and/or hidden double-mode pulsation is behind  the excitation of this peculiar frequency component.

\section{Period doubling}\label{pdsec}

\begin{figure}
\includegraphics[width=9.1 cm]{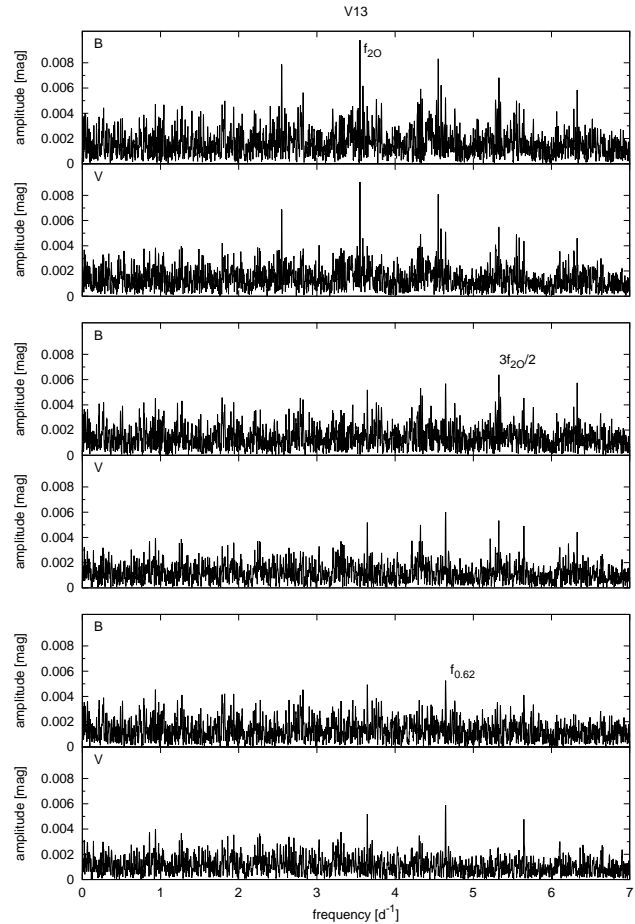}
\caption{ $B$ and $V$ residual spectra of V13 in consecutive prewhitening steps.
The largest signal appears at the $f_{\mathrm{2O}}$ (2O-mode) frequency in the data prewhitened for the $f_{\mathrm{F}}$ and $f_{\mathrm{1O}}$ frequency components, their modulations and the linear-combination terms  (top panels).
In the next prewhitening steps, frequencies are detected at $3f_{\mathrm{2O}}/2$, and then, at $f_{0.61}$  (middle and bottom panes).\label{13pd}}
\end{figure}

\begin{figure}
\includegraphics[width=9.1 cm]{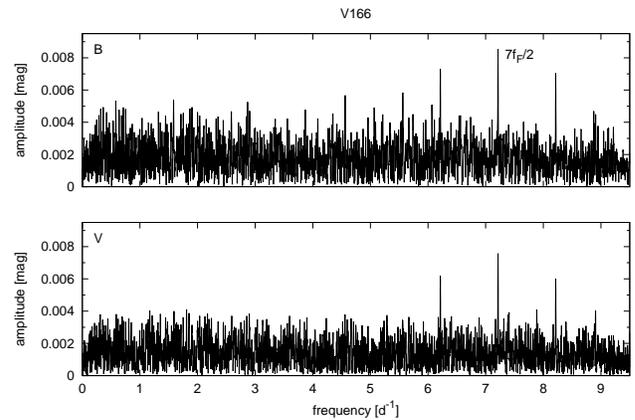}
\caption{ $B$  and $V$ residual spectra of V166, prewhitened for the $f_{\mathrm{F}}$ and $f_{\mathrm{1O}}$ frequency components and their modulations and linear-combinations, are drawn. Frequency appears at the position of the half-integer frequency,  $7f_{\mathrm{F}}/2$. \label{166pd}}
\end{figure}

A new feature of the pulsation of RRab stars, the alternating shapes of their light curves (i.e., period doubling)  was recognized, thanks to the high-precision  time-resolved photometry of the Kepler space-telescope data \citep[as a review see][]{sz14}. Half-integer frequencies of the fundamental-mode frequency are the signature of the phenomenon in the Fourier spectrum.
The observed half-integer frequencies  had the highest amplitude typically at $3f_{\mathrm{F}}/2$ and they were detected exclusively in stars showing Blazhko modulation in the Kepler data. \cite{mo15}  reported that  half-integer frequencies also appear in RRc stars, however, in this case the period doubling was  connected to the additional,  $f_{0.61}$ frequency, instead of the radial-mode pulsation component. Period doubling of the third (third radial overtone or non-radial) mode of OGLE-BLG-RRLYR-24137 was also reported by \cite{smo15b}. 
Using turbulent, convective hydrodynamical models and amplitude equation formalism, \cite{k11} and \cite{kb}  argued that a 9:2 resonance between the fundamental and the ninth radial overtone (which is a strange mode) is responsible for the observed period doubling of RRab stars and it might be taken part in the generation of the Blazhko modulation.  In contrast to this claim, citing \cite{mo15} `it is unclear which (if any) resonance is responsible for the period doubling in the RRc and RRd variables.'

The D12 data revealed period doubling of two multi-mode RR Lyrae stars in M3, as well.
It is connected to the second overtone of the triple-mode star, V13, and  to the fundamental mode of the double-mode star, V166. Both stars show Blazhko modulation of the fundamental and the overtone modes.
Table~\ref{pd} gives the main properties of the frequencies connected to period doubling  of these stars. These frequencies appear at the exact value of the  half-integer frequencies  within the uncertainty limits ($3\sigma$). 

The  $B$  and $V$ residual spectra, prewhitened for the $f_{\mathrm{F}}$ and $f_{\mathrm{1O}}$ frequency components, their modulations and the linear-combination terms, are shown for V13 and V166 in Fig.~\ref{13pd} and Fig.~\ref{166pd}, respectively. The frequency of the 2O appears first  in V13, then, in further prewhitening steps, frequencies are detected at $3f_{\mathrm{2O}}/2$, and finally, at $f_{0.61}$. The prewhitened spectra of V166 show a signal at the   $7f_{\mathrm{F}}/2$ half-integer frequency with a high significance.
 
Despite the fact that the amplitude of the 2O pulsation mode is only 0.010~mag, quite surprising that the half-integer frequency is connected to the 2O mode and not to the most strongly modulated large-amplitude fundamental mode. The amplitude ratio of  the half-integer frequency to the 2O is quite large, $A_{3f_{\mathrm{2O}}/2}/A_{f_{\mathrm{2O}}}=0.50$. We note that larger amplitude ratio of any half-integer component was detected only in  some of the $f_{0.61}$ frequencies reported by \cite{mo15} and \cite{netzel}. The half-integer frequency was connected also to a small-amplitude frequency component and not to the large-amplitude radial mode in these cases.

\begin{table}
\begin{center}
\caption{Period doubling detected in the M3 data.\label{pd}}
\begin{tabular}{l@{\hspace{0mm}}c@{\hspace{2mm}}c@{\hspace{2mm}}c@{\hspace{3mm}}c@{\hspace{3mm}}c@{\hspace{3mm}}l}
\hline
\hline
Star &Mode\tablenotemark{a}& $k$\tablenotemark{b} & $P$ [d]& Error & $P_{\mathrm {PD}}-2P/k$    & $A(V)$ \\
\hline
V13   &2O & & 0.28160 & 0.00003  &         & 0.010    \\
            &&3   & 0.18775 & 0.00003  & 0.00002 & 0.005 \\

V166&F&   & 0.48504 & 0.00001  &         & 0.200     \\
            &&7   & 0.13861 & 0.00002  & 0.00003 & 0.008 \\
\hline
\end{tabular}
\tablenotetext{1}{Pulsation mode period-doubling belongs to.}
\tablenotetext{2}{Order ($k$) of the half-integer frequency, $2P/k$.}
\end{center}
\end{table}

Looking for a candidate resonance behind the period doubling of the 2O mode of V13, a low-order one,  $3f_{\mathrm{2O}}=2f_{\mathrm{4O}}$, may be valid in the $6800-6900$~K temperature range based on the period-ratio grid of radial modes \citep[fig. 3 in][]{k11}. According to the temperature calibration of \cite{ca05}, double-mode stars   fall exactly into this temperature range in M3. However, as the fourth overtone is strongly damped \citep[see fig. 4 in][]{k11}, its excitation, even in resonance position is dubious. Why period doubling of the small-amplitude 2O appears in V13 and whether its $3f_{\mathrm{2O}}/2$ frequency can be be explained by a resonance with some high-order overtone remain a question to be answered. Besides the period doubling feature observed in RRab stars what V166 seems to be following, the behavior of V13 is pointing to some additional mechanism also taking part in this effect.

 Period doubling of the overtone mode of single or double-mode stars or the  $f_{0.61}$ frequency component is not detected in M3. 

\section{Blazhko effect}\label{blsec}

\begin{table*}
\begin{center}
\caption{Blazhko properties according to the D12 data.\label{bl.tab} }
\begin{tabular}{l@{\hspace{2mm}}c@{\hspace{2mm}}c@{\hspace{2mm}}c@{\hspace{2mm}}c@{\hspace{1mm}}c@{\hspace{1mm}}l@{\hspace{-1mm}}}
\hline
\hline
Star&  $P_{\mathrm{puls}}$  & $P_{\mathrm{mod}}$\tablenotemark{a} &  $f_{\mathrm{mod}}$ &$A_{\mathrm{A_{\mathrm{mod}}}}$\tablenotemark{b}  &$A_{\mathrm{ph_{\mathrm{mod}}}}$\tablenotemark{c} &Summary of the main modulation components\\
&  d  & d&d$^{-1}$& mag & rad &\\
\hline
V13 & 0.35072 & 139(1)& 0.007  & 0.20& 1.25&  $-f_{\mathrm m}$, $-2f_{\mathrm m}$ at the first and 2nd harmonic orders \\
    & 0.47949 & 139(1)& 0.007  & 0.42& 0.85&  dominant $-f_{\mathrm m}$, some small amplitude $+f_{\mathrm m}$, $-2f_{\mathrm m}$  in the first 4 harmonic orders\\
V44 & 0.36812 &  56.0(8)& 0.018 & 0.15& 2.20&  complex multiplets at the 1st and 2nd harmonic orders\\
    & 0.50377 &  97.0(2)& 0.010 & 0.45& 1.00&  complex modulation with strong $-f_{\mathrm m}$, $-2f_{\mathrm m}$ and $\pm f_{\mathrm m}/2$ components \\
V70 & 0.48607 &  150(10)&0.007 & 0.01& 0.10&  triplet with 0.007 and 0.006 mag amplitude of the  $-f_{\mathrm m}$, $+f_{\mathrm m}$ components\\
V99 & 0.3611 &  40.3(2)&0.025 & 0.35& 2.20&  $+f_{\mathrm m}$, $+2f_{\mathrm m}$, $+3f_{\mathrm m}$ components at the 1st and 2nd harmonic orders \\
    & 0.4821 & $\approx450$&$\approx0.002$\,\,\,\,\,  &$>0.25$\,\,\,\, &$>0.2$\,\,\,&  triplet with 0.050 and 0.025 mag amplitude of the  $-f_{\mathrm m}$ and $+f_{\mathrm m}$ components\\ 
V126& 0.34841 &  15.0(2)&0.067 & 0.03& 0.04&   one component at $f+f_{\mathrm m}$ with 0.006 mag amplitude\\
V140& 0.33316 &  28.8(1)&0.035 & 0.10& 0.63&  triplets at least up to the 5th harmonic order \\
V166& 0.35672 &  43.5(5)&0.023 & 0.08& 0.50&  one component at $f-f_{\mathrm m}$ with 0.023 mag  amplitude\\
    & 0.48504 &  71.5(3)&0.014 & 0.45& 1.20&   members of the quintuplets (dominant $-f_{\mathrm m}$, $-2f_{\mathrm m}$) up to the 4th harmonic order\\
V168& 0.27594 &  23.0(1)& 0.043& 0.16& 1.00&  members of the quintuplets  up to the 4th harmonic order   \\
V203& 0.28979 & 118(5)& 0.008  & 0.03& 0.15&   one component at $f+f_{\mathrm m}$ with 0.007 mag amplitude\\
\hline
\end{tabular}
\tablenotetext{1}{The error in the last digit is given in parenthesis.}
\tablenotetext{2}{Full amplitude of the amplitude modulation in $V$ band.}
\tablenotetext{3}{Full amplitude of the phase modulation in $V$ band.}
\end{center}
\end{table*}

\begin{figure*}
\includegraphics[width=17.5 cm]{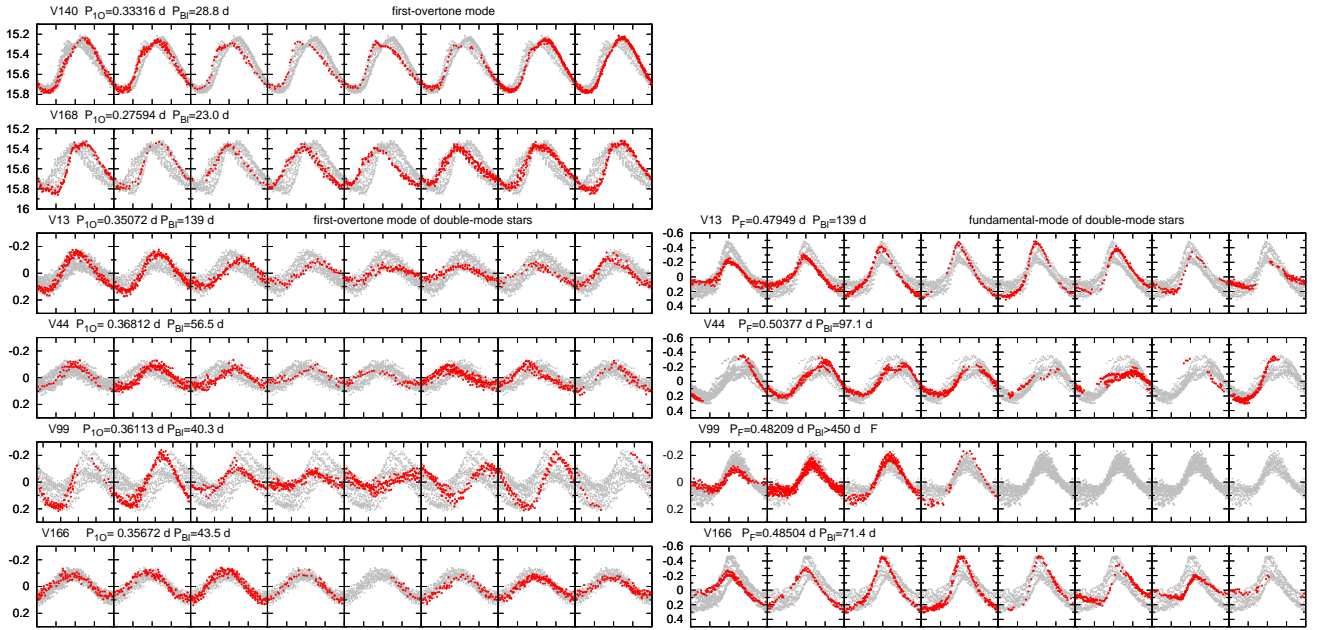}
\caption{$V$ light curves of the Blazhko stars in eight phases of their main modulation cycle are shown. The light curves of the double-mode stars are prewhitened for the light variation of the other mode and for the linear-combination terms.\label{blfig}}
\end{figure*}

Equally spaced asymmetric triplets, multiplets, which are characteristics for Blazhko RRab stars,  are detected in the residual spectra of seven overtone and double-mode stars. A  small-amplitude  component near the radial-mode frequency appears in the residual spectra of two further  stars (V126, V203). Although the non-radial mode origin of these latter signals cannot be excluded, we classify these stars also as Blazhko variables.  Six of the Blazhko stars are multiperiodic besides their light-curve modulation. V13, V44, V99, and V166 are double(triple)-mode stars, and the $f_{0.61}$ frequency component is detected in V70, and V126. Some specialties of the Blazhko properties of overtone stars in M3 were already discussed in \cite{rrdbl}. The Blazhko stars are denoted by thick symbols in Figs.~\ref{p-va} and \ref{four}. 


Table \ref{bl.tab} lists the pulsation and modulation periods, the amplitudes of the amplitude- and phase-modulations, and  data and comments on the main features of the modulation detected in the D12 of Blazhko stars.

The light curves of the strongly modulated stars  in eight phases of the Blazhko cycle are shown in Fig.~\ref{blfig}. The data of multi-mode pulsators are prewhitened for the  signals of the pulsation and the modulation of the other radial mode, the $f_{0.61}$ frequency, and for the linear-combination terms. Note that the modulation periods of the two modes are different in V44, V99, and V166.

The light curves in Fig.~\ref{blfig}, and the data given in Table \ref{bl.tab} show that the modulation of RRc and  the overtone-mode of RRd stars are  phase modulations predominantly. The phase modulation of the overtone mode of V44 and V99 is larger than one third of the pulsation period. This is in contrast with the Blazhko behavior of the fundamental-mode of RRd  and  RRab stars. Their modulation is dominated mostly by amplitude variation or  their amplitude and phase modulations have  similar strength.

Pre-maximum bump appears only in the two single-mode Blazhko stars (V140 and V168) in some phases of their Blazhko cycle.  Both of these stars are at high luminosity but at different temperatures as indicated by their magnitudes,  periods, and color indices (see Fig.~\ref{p-va}).  The bump is observed at the opposite phases of the modulation of the two stars. The shock-induced phenomenon appears in V140 and V168 when the phase-shift  is negative (period is decreasing) and the amplitude is small  and when the phase-shift of the light curve is positive (period is increasing) and the amplitude is large. The temperature difference between the two stars may explain this reverse behavior. The temporal appearance of the bump in Blazhko RRc stars implies that their main physical properties vary during the Blazhko cycle, similarly as detected in Blazhko RRab stars \citep[see e.g.][]{mw2,czl,rzl}.


\subsection{Notes on the Blazhko properties of the stars}\label{blnotesec}

Besides the analysis of the D12 data, the archive  photographic and CCD data collected in J12 have also been checked to follow the changes of the Blazhko properties on a long time scale. Further details on the pulsation/modulation of Blazhko RRd stars are given in Section~\ref{dmsec}.

{\bf V13} is a triple-mode star (F/1O/2O) with a dominant fundamental mode (Sect.~\ref{v13sec}). The F and 1O modes are modulated with the same period and the amplitude and phase variations of the modes are in anti-phase with each other \citep[see also in][]{rrdbl}, similarly to 1O/2O Blazhko Cepheids \citep{mk09}.  The amplitude ratios of the radial modes  to the largest-amplitude modulation components are 2.3 and 1.6 for the F and 1O modes.  Assuming that the  pulsation/modulation amplitude ratios of each radial mode are similar, the modulation amplitudes are expected around $0.004-0.005$~mag at the 2O frequency as the $V$ amplitude of the 2O mode is 0.009~mag. However, no modulation component with amplitude larger than $\sim$$0.002$~mag has been found; the largest amplitude signal close to the 2O frequency in the residual spectrum is the 1~d$^{-1}$ alias component of the  $f_{0.61}$ frequency (see Fig.~\ref{13pd}). 
The previous CCD observations show a similar behavior as in the D12 data; anti-phase modulations of the fundamental and overtone modes with 144~d is derived from  the combined K98, B06 and J12 data set.   \\
{\bf V44} is a double-mode star with a dominant fundamental mode. The modulation of the fundamental mode is very complex; the main modulation period is 97~d but modulations with the half and the twice of this period are also detected. The 56~d modulation of the overtone mode is dominantly phase modulation. Modulation of the fundamental mode with 52~d, 104~d, and 111~d are identified in the C01, H05+B06, and the J12 data, respectively. The photographic data show some signs of modulation with $P_{\mathrm m}=82$~d (JD~$2420625-2424684$), and 49~d (JD~$2428963-2436991$)
 \\
{\bf V70} is a very-long-period (0.486~d), over-luminous, evolved variable. The small-amplitude, 150-d  modulation detected in the D12 data is dominantly phase modulation but amplitude changes  are also observed mostly in the $B$ band. The amplitude changes and the 150-d period of the modulation, which is shorter than the time span of the observations, make it unlikely that period change mimics Blazhko modulation in this case. The C01, H05, B06, and J12 data show significant residuals at  $f_{\mathrm{1O}}$  but with different separations. However, it is not clear in these cases whether these signals arise from strong period changes, which  are typical in this star,  or from light-curve modulation.\\
{\bf V99} is a double-mode star. The amplitude of the fundamental mode was increasing from 0.13 to 0.42~mag in the D12 data.  The $V$ amplitude of the fundamental mode  was 0.20, 0.40 and 0.25~mag  in $1992-1993$, 1999 and 2009, respectively, according to the previous CCD observations (C01, H05, B06, and J12). The strong differences in the measured amplitudes of the fundamental mode indicates that the amplitude increase of the fundamental mode in the D12 data reflects Blazhko modulation on a time-scale of some hundred days and that the true amplitude of the fundamental mode can be even larger as measured in the D12 data.  The overtone mode is strongly amplitude and phase modulated. Blazhko modulation of the overtone mode is identified in both the combined H05 and B06  and the J12 data with a period of  around 40~d.\\
{\bf V126} is an RRc star showing a small-amplitude $f_{\mathrm{1O}}+f_{\mathrm m}$ component. This  RRc star is the longest period, not over-luminous one in the magnitude calibrated sample, which lies just at the blue edge of the double-mode region.\\
{\bf V140} is about 0.2~mag above the ZAHB. The J12 data show Blazhko modulation ($f_{\mathrm{1O}}+f_{\mathrm m}$, $f_{\mathrm{1O}}-f_{\mathrm m}$ and $2f_{\mathrm{1O}}-f_{\mathrm m}$ components) with $f_{\mathrm m}=0.0689$~d$^{-1}$ ($P_{\mathrm{m}}=14.5$~d), which is half of the recent modulation period.
The combined H05 and B06 observations  yield similar light-curve modulation as observed in the D12 data. The pulsation and modulation periods were 0.333146~d and 29.0~d in 1999. The modulation period cannot be determined in the C01 data, but from the large residuals around the pulsation frequency we suppose that the light curve was modulated between JD~2448755 and 2449091 also. Similar modulation as observed in the CCD data is suspected in two parts of the photographic data; the $f_{\mathrm{1O}}+0.068$~d$^{-1}$ ($\sim$$15$~d) component  might be present in the data between JD~$2422729-2424684$ and JD~$2433390-2435933$, in the latter data set the $f_{\mathrm{1O}}-0.034$~d$^{-1}$ ($\sim$$29$~d) component has a significant amplitude, too. The dominant modulation of V140 seems to switch between $f_{\mathrm m}$ and $2f_{\mathrm m}$ time to time.\\
{\bf V166} is a double-mode star with a dominant fundamental mode. The fundamental mode shows strong amplitude and phase modulations, while the modulation of the overtone mode is modest. A 71~d modulation of the fundamental mode is also detected in the B06 and J12 data, and the large residual at $f_{\mathrm{F}}$ points to modulation of the fundamental mode in the C01 data, too.\\
{\bf V168} is a short-period, over-luminous star. The combined H05 and B06 data   yield similar light-curve modulation as observed in the D12 data. The pulsation and modulation periods were  0.27595~d and 22.9~d in 1999. Modulation of the fundamental mode is also evident in the C01 data. \\
{\bf V203} is a second-overtone over-luminous variable. A frequency component at $f_{\mathrm{2O}}+0.0096$~d$^{-1}$  is also detected in the J12 data with 0.008 mag amplitude. 

\section{Double- and triple-mode stars}\label{dmsec}

\begin{figure}
\includegraphics[width=9.2 cm]{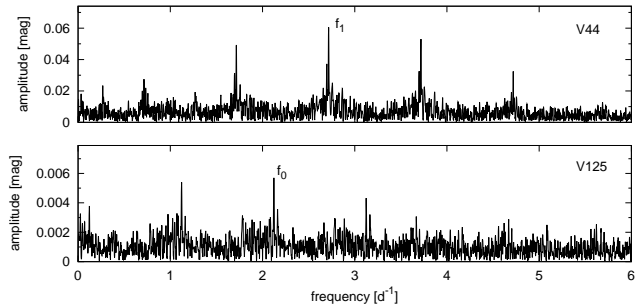}
\caption{Residual spectra of the detected new radial-mode components of V44 and V125. The data are prewhitened for all the other signals.\label{ressp}}
\end{figure}

Nine double-mode stars have been identified in M3 previously. A thorough investigation of M3 RRd stars was given by \cite[][hereafter C04]{cc04}. Double-mode nature of  two more  stars,  V44 and V125, and the triple-mode behavior of V13 are detected in the D12 data.   The residual spectra prewhitened for all the other signals in Fig.~\ref{ressp} depict the new radial modes of V44 and V125 in the D12 data like the 2O mode of V13 in Fig.~\ref{13pd}.

The D12 data makes it possible to reveal the complex behavior of RRd stars.  To inspect the changes of pulsation properties all the previous photometric data listed in Section~\ref{datasec} are also checked. 
The 200-d length of the D12 data allows to determine the pulsation frequencies of double-mode stars typically with a precision of the order of $10^{-4}-10^{-5}$~d$^{-1}$.
The frequency spectra of the previous photometric data  are strongly biased; especially the 1~yr$^{-1}$ alias frequencies result in $\pm0.003$~d$^{-1}$ ambiguities of the solutions.
Therefore, the frequencies published in the literature (K98, C01, C04, H05, B06, and J12) were revised in the sense that from the equally possible frequency solutions 
that was accepted which gave closest resemblance to the solution of the D12 data.


The main photometric properties of the double/triple mode stars obtained from the D12 data are given in Table~\ref{rrd.dat}. Table~\ref{dm.tab} summarizes the detected pulsation and modulation (discussed in Section~\ref{blsec}) properties of double-mode stars at different epochs.


\begin{table*}
\begin{center}
\caption{Long-term changes in the properties of double-mode stars according to a reanalysis of the photometric data.\label{dm.tab}}
\scriptsize
\begin{tabular}{@{\hspace{0mm}}l@{\hspace{1mm}}l@{\hspace{1mm}}c@{\hspace{2mm}}l@{\hspace{2mm}}c@{\hspace{2mm}}l@{\hspace{2mm}}l@{\hspace{1mm}}c@{\hspace{0mm}}l@{\hspace{1mm}}l@{\hspace{1mm}}l@{\hspace{1mm}}l}
\hline
\hline
Star& Band & JD-2400000&  $f_{\mathrm{F}}$ & ${A_{\mathrm{F}}}$\tablenotemark{a}   & $f_{\mathrm{1O}}$ & $A_{{\mathrm{1O}}}$   & $A_{{\mathrm{F}}}/A_{{\mathrm{1O}}}$ &$f_{\mathrm{F}}/f_{\mathrm{1O}}$ & ${f_{\mathrm{m_{\mathrm{F}}}}}$\tablenotemark{b} & ${f_{\mathrm{m_{\mathrm{1O}}}}}$\tablenotemark{c}     & Remark \\
&&& d$^{-1}$&mag&d$^{-1}$&mag&&&d$^{-1}$&d$^{-1}$&\\
\hline
V13 &pg    & 23858-24684&  2.07019& 0.46 &          &      &      &        &           &                &     \\
    &pg    & 28963-35933&  2.07022& 0.36 &          &      &      &        &           &                &     \\
    &pg    & 36367-37131&  2.07023& 0.38 &          &      &      &        &           &                &     \\
    &pg    & 39944-47554&  2.08544& 0.28 &          &      &      &        &           &                &    \\
    &CCD $V$ & 50162-54965&  2.08555& 0.22 &   2.85123& 0.07 &  3.3 & 0.7315 & 0.0069    &  0.0069        &   \\
    &CCD $V$ & 55935-56137&  2.08555& 0.23 &   2.85127& 0.08 &  2.9 & 0.7314 & 0.0072    &  0.0072        &$f_{\mathrm{2O}}$, $f_{PD}=3f_{\mathrm{2O}}/2$, $f_{0.61}$  \\
V44 &pg    & 13372-22840&  1.97485& 0.58 &          &      &      &        &           &                &     \\ 
    &pg    & 24283-30078&  1.97490& 0.54 &          &      &      &        &           &                &     \\ 
    &pg    & 31965-36991&  1.97471& 0.45 &          &      &      &        &           &                &     \\  
    &pg    & 37018-47555&  1.97470& 0.30 &          &      &      &        &           &                &     \\   
    &CCD $V$ & 48755-49091&  1.975:& 0.22 &   2.7:   & 0.06 &  3.7 & 0.731: &  0.019:   &    & \\
    &CCD $V$ & 50893-51257&  1.9744 & 0.40 &          &      &      &        &0.010/0.019&    &               \\
    &CCD $V$ & 54869-54965&  1.9736 & 0.30 &   2.685: & 0.03 &  10.0 & 0.7350&0.009/0.018&    &  \\
    &CCD $V$ & 55935-56137&  1.98505& 0.21 &   2.71650& 0.06 &  3.5 & 0.7307 &0.005/0.010/0.020&0.018 &\\
V68 &pg    & 22455-24684&  2.08989& 0.25 &   2.80916& 0.26 &  0.8 & 0.7440 &           &                &     \\
    &pg    & 33034-39944&  2.08985& 0.22 &   2.80920& 0.25 &  0.8 & 0.7439 &           &                &     \\
    &CCD $V$ & 48755-50176&  2.08985& 0.17 &   2.80921& 0.20 &  0.8 & 0.7439 &           &                &     \\
    &CCD $V$ & 50893-51257&  2.08988& 0.17 &   2.80918& 0.19 &  0.9 & 0.7439 &           &                &     \\
    &CCD $V$ & 54869-54965&  2.0899 & 0.17 &   2.8092 & 0.21 &  0.8 & 0.7440 &           &                &     \\
    &CCD $V$ & 55935-56137&  2.08986& 0.17 &   2.80921& 0.20 &  0.9 & 0.7439 &           &                &   $f_{0.61}$    \\
V79 &pg    & 13372-20656&  2.0666& 0.66 &           &      &      &        &           &                &     \\
    &pg    & 22729-24684&  2.0697& 0.51 &           &      &      &        &           &                &     \\
    &pg    & 28963-35933&  2.0690& 0.50 &           &      &      &        &  0.050:   &                &     \\
    &pg    & 36367-39672&  2.0691& 0.44 &           &      &      &        &  0.012:   &                &   \\
    &pg    & 41061-42970&  2.0698& 0.25 &           &      &      &        &           &                &  residual at $f_{\mathrm F}$\\
    &pg    & 43160-44760&  2.0691& 0.42 &           &      &      &        &           &                &  residual at $f_{\mathrm F}$\\
    &pg    & 44988-46566&  2.0696& 0.43 &           &      &      &        &           &                &  \\
    &pg    & 47202-47329&  2.0711& 0.49 &           &      &      &        &  0.015:  &                &  signal at 2.087 d$^{-1}$  \\
    &pg    & 47552-47686&        &      &          &      &      &        &           &                &  at JD 2447620 pulsation ceases 
 \\
    &pg    & 47970-48485&  2.0843 & 0.29 &         &      &      &        &           &                & strong residual at 2.0743 d$^{-1}$ \\
    &CCD $V$ & 48755 49091&  2.0856:& 0.11 &  2.7931:& 0.16 &  0.7 & 0.7467 &           &                &     \\
    &CCD $V$ & 50162-50176&  2.084  & 0.09 &  2.792  & 0.17 &  0.5 & 0.746  &           &                & \\
    &CCD $V$ & 50893-50972&  2.0865 & 0.09 &  2.7914 & 0.17 &  0.5 & 0.7474 &  0.015    &                &   \\
 &CCD clear& 54257-54273&  2.072  & 0.32  &  2.795 & 0.13 &  2.5 & 0.741  &           &                &  \\
    &CCD $V$ & 54869-54965&  2.06913& 0.41 &         &      &      &        &  0.016 && largest mod. peak at  2.085 d$^{-1}$ \\
    &CCD $V$ & 55935-56137&  2.06914& 0.35 &         &      &      &        &  0.006/0.016/0.055 & & 3rd largest mod. peak at  2.085 d$^{-1}$  
\\
V87 &pg    & 22729-24684& 2.08184&  0.14&  2.79737 & 0.32 &  0.5 & 0.7442 &           &                &     \\
    &pg    & 33390-39944& 2.08261&  0.09&  2.79732 & 0.25 &  0.4 & 0.7445 &           &                &     \\
    &pg    & 42838-47555& 2.08147&  0.13&  2.79733 & 0.29 &  0.5 & 0.7441 &           &                &     \\
    &CCD $V$ & 48755-50176& 2.08248&  0.08&  2.79732 & 0.20 &  0.4 & 0.7445 &           &                &     \\
    &CCD $V$ & 50893-51283& 2.08238&  0.09&  2.79738 & 0.20 &  0.5 & 0.7444 &           &                &     \\
    &CCD $V$ & 54869-54965& 2.08263&  0.09&  2.79715 & 0.22 &  0.4 & 0.7446 &           &                &     \\
    &CCD $V$ & 55935-56137& 2.08260&  0.08&  2.79729 & 0.19 &  0.4 & 0.7445 &           &  &$f_{0.61}$    \\
V99 &pg    & 13372-15161&        &      &  2.77062 & 0.42 &      &        &     &         &     \\
    &pg    & 22455-24684& 2.0673:&  0.22&  2.7712: & 0.23 &  0.8 & 0.7460:&     &         &     \\
    &pg    & 33034-35933& 2.0719 &  0.10&  2.7712  & 0.21 &  0.5 & 0.7476 &     &         &     \\
    &pg    & 36367-39933& 2.0732 &  0.14&  2.7711  & 0.20 &  0.6 & 0.7482 &     &         &   residual at 2.069        \\
    &pg    & 42838-43254&        &      &  2.7711  & 0.20 &      &        &     &         &     \\
    &CCD $V$ & 48755-49091& 2.0746:&  0.08&  2.7707: & 0.24 &  0.3 & 0.7488: &           &                &     \\
    &CCD $V$ & 50893-50972& 2.0740 &  0.15&  2.7696  & 0.10 &  1.4 & 0.7488 &           &   0.025        &     \\
    &CCD $V$ & 54869-54965& 2.0734 &  0.10&  2.7703  & 0.18 &  0.6 & 0.7484 &           &   0.025        &     \\
    &CCD $V$ & 55935-56137& 2.0743 &  0.12&  2.7691 & 0.09 &  1.4 & 0.7490 &  0.002:   &   0.025        &     \\ 
V125&pg    & 13372-24684&         &      & 2.85862 & 0.36 &      &        &           &                &     \\ 
    &pg    & 28963-39944&         &      & 2.85860 & 0.28 &      &        &           &                &     \\ 
    &pg    & 42838-47555&         &      & 2.85859 & 0.26 &      &        &           &                &     \\ 
    &CCD $V$ & 48755-50176&         &      & 2.85860 & 0.21 &      &        &           &                &     \\ 
    &CCD $V$ & 50893-50972&         &      & 2.85869 & 0.20 &      &        &           &    & \\
    &CCD $V$ & 54869-54965&         &      & 2.85857 & 0.21 &      &        &           &    & \\
    &CCD $V$ & 55935-56137& 2.1233  & 0.01 & 2.85862 & 0.21 & 0.03 & 0.7428 &           &    & $f_{0.61}$ \\ 
V166&CCD $V$ & 48755-50176 & 2.0619 & 0.18 & 2.8031  & 0.14 & 1.2 & 0.7356 &            &        & strong residual at $f_{\mathrm{F}}$\\
    &CCD $V$ & 50893-51283 & 2.0621 & 0.20 & 2.8031  & 0.09 & 2.3 & 0.7356 & 0.014      &        & \\
    &CCD $V$ & 54869-54965 & 2.0623 & 0.23 & 2.8034  & 0.12 & 1.9 & 0.7356 & 0.014      &        & \\
    &CCD $V$ & 55935-56137 & 2.0617 & 0.20 & 2.8033  & 0.09 & 2.2 & 0.7355 & 0.014      &  0.023 & $f_{PD}=7f_{\mathrm{F}}/2$\\
 V200&CCD $V$ & 50893-51283 & 2.0606 & 0.19 & 2.7700  & 0.18 & 1.0 & 0.7439 &            &        &      \\
&CCD flux ($V$)& 50920-50965 & 2.0606 &      & 2.7690  &      & 0.8 & 0.7442 &            &        &      \\
&CCD flux ($B$)& 55935-56137 & 2.06058&      & 2.76995 &      & 0.9 & 0.7439 &            &        &      \\
V251   &CCD $V$ & 50896-51283 & 2.0998:& 0.18 & 2.8232: & 0.15 & 1.2 & 0.7438 &           &        &      \\ 
&CCD flux ($B$)& 55935-56137 & 2.0998&      & 2.8230 &      & 1.0 & 0.7438 &           &        &      \\
V252&CCD $B$ & 48755-50622 & 2.2071:& 0.28 & 2.9754: & 0.32 & 0.9 & 0.7418 &           &        &    \\
    &CCD $V$ & 50893-51283 & 2.2076& 0.18 & 2.9755 & 0.19 & 0.9 & 0.7419 &           &        &      \\
&CCD flux ($B$)& 55935-56137 & 2.2075&      & 2.9754 &      & 1.0 & 0.7419 &           &        &      \\
\hline
\end{tabular}
\vspace{-1em}
\tablenotetext{1}{Fourier amplitude;  $A_{pg}$  is about 1.3 times larger than $A_V$, at the earliest epochs (JD 2413000-2415000) it is about 2 times larger.}
\tablenotetext{2}{Modulation frequency of the F mode.}
\tablenotetext{3}{Modulation frequency of the 1O mode.}
\end{center}
\end{table*}

\subsection{The triple mode star, V13}\label{v13sec}

{\bf{V13}} The new frequency components detected  in the residual spectra of V13 are shown in Fig~\ref{13pd}.  The  frequency appearing  at $f_{\mathrm{F}}/f_{\mathrm{x}}=0.5873$ ($f_{\mathrm{1O}}/f_{\mathrm{x}}= 0.8028$) frequency ratio matches the required  $P_2/P_0$ period-ratio range ($0.582-0.593$) of F$+$2O double-mode RR Lyrae-type stars \citep{mo14}. Therefore, we identify this signal with the second radial overtone mode. The radial-mode frequencies are far from the $2f_{\mathrm{1O}}=f_{\mathrm{F}}+f_{\mathrm{2O}}$ resonance condition,  $2f_{\mathrm{1O}}-f_{\mathrm{2O}} = 2.1508$, while $f_{\mathrm{1O}}$ is observed at 2.0855~d$^{-1}$.

Modeling the radial-mode frequencies of triple-mode stars provides a unique opportunity to determine the physical parameters of the stars precisely.
Unfortunately, pulsation models fail to explain until yet the anomalous period ratios observed in M3 and in the Galactic Bulge \citep[C04,][]{smo15a}, and V13 is one of the RRd stars with a too small period ratio. According to the D12, and the combined K98, B06 and J12 data, its $f_{\mathrm{1O}}/f_{\mathrm{F}}=0.7315$ frequency ratio is even smaller than  that was derived by C04 (0.7379). 

V13 was identified first as a double mode variable by C04. It had been classified as an RRab star in all the previous studies. The fundamental mode was the dominant mode in each available data set. 

The pulsation period decreased drastically by 0.0036~d ($\Delta f_{\mathrm{F}}=0.015$~d$^{-1}$) in the 1960s. The period change was accompanied by a 25\% decrease of the fundamental-mode amplitude. Unfortunately, the quality of the photographic data does not make it possible to detect small-amplitude signals in these data. Therefore, it cannot be excluded that the strong period change  and amplitude decrease were synchronous with the appearance of the overtone mode. 

\begin{figure}
\includegraphics[width=8.2 cm]{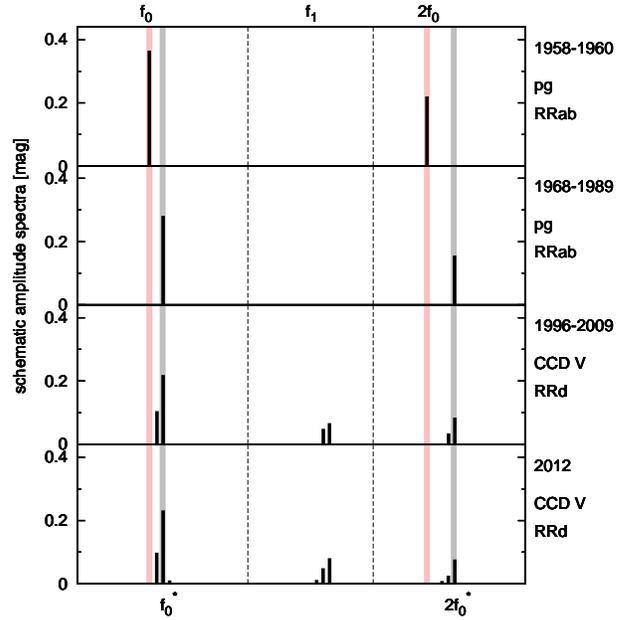}
\caption{Evolution of the $f_{\mathrm{F}}$, $2f_{\mathrm{F}}$, and $f_{\mathrm{1O}}$ frequency components of V13 in the $1958-2012$ period. Pink and gray strips indicate the positions of the fundamental mode and its first harmonic detected at the different epochs.\label{13}}
\end{figure}

The reanalysis of all the CCD data shows that, in fact,  the overtone mode is present in all  these observations.  The 0.015~d$^{-1}$ frequency shift of the fundamental mode preceding the appearance of the 1O mode is about twice of the modulation frequency (0.0072~d$^{-1}$) detected in the CCD data.  The $f_{\mathrm{F}}$, $2f_{\mathrm{F}}$, and $f_{\mathrm{1O}}$ frequencies identified in the different data sets of V13 are shown schematically in Fig.~\ref{13}. 

\begin{figure}
\includegraphics[width=8.2 cm]{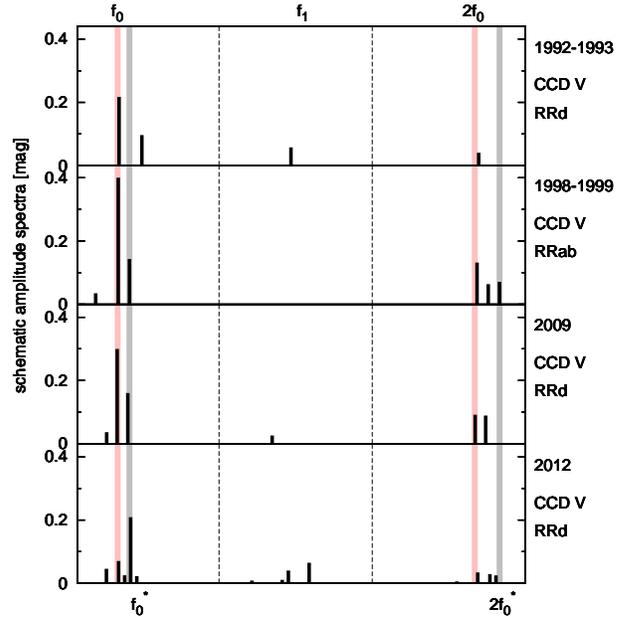}
\caption{The same as Fig.~\ref{13} for the frequency components of V44 in the $1998-2012$ period.\label{44}}
\end{figure}

\subsection{The double-mode stars}\label{dmnotesec}

{\bf{V44}}  was classified as an RRab star with a complex light-curve variation assigned to a very strong Blazhko effect previously. Its  0.506-d fundamental-mode period is significantly longer than the $0.453-0.485$-d period of the other RRd stars. There are 18 RRab stars in M3 that  do not show the overtone mode, but their periods are shorter than the F-mode period of V44. 

We find, however,  that frequency around the possible position of the 1O-mode appears already in the C01 and J12 observations, not only in the D12 data. It is the highest amplitude signal in the residual spectra of these data sets, but 1O is not detected in the H05 and B06 observations. The amplitude of the fundamental mode is $1.5-2.0$ times larger in the H05 and B06 data than in those observations when the overtone mode is also present, though with a small amplitude. 

The detected frequencies of V44 at the positions of $f_{\mathrm{F}}$, $2f_{\mathrm{F}}$ and $f_{\mathrm{1O}}$ are shown for the available CCD data sets schematically in Fig.~\ref{44}.
The frequency of the main component at $f_{\mathrm{F}}$ is 0.011~d$^{-1}$ larger (the period is 0.0029~d shorter) in the D12 data than it was previously. The frequency, which is identified as $f_{\mathrm{F}}-f_{m}$ in the D12 data is the largest-amplitude signal in the 2009 observations and in all the other CCD data. Despite of the fundamental-mode frequency having a significantly larger amplitude than the largest side frequency (the amplitude ratio is $\sim$$3$) in the D12 data,  $2f_0$ and $3f_0$ are not detected;  only modulation components appear in the 2nd, and 3rd harmonic orders. In contrast, harmonic components of the F-mode frequency are detected up to harmonic orders $4-5$ in  other CCD data.

The  frequency ratio of the radial modes has an anomalously low value in each data showing double-mode pulsation. \\
{\bf{V68}} All the CCD and photographic data yield a very similar solution with $0.7439-0.7440$ frequency ratio and with a dominant overtone mode. \\
{\bf{V79}} Its unique behavior was discussed in several previous papers \citep[e.g.][]{cl97,cl99,v79ct,go10}. Data published in \cite{cl99} and \cite{gb07} are also utilized in the analysis besides the other CCD and photographic data.

V79 was a double mode star only in the $1992-2007$ interval.  Before and after this double-mode episode it had/has been an RRab star. A strong Blazhko effect has been detected since 2008. Indications of Blazhko modulation of the fundamental mode are evident in many parts of the photographic data, but no Blazhko period can be derived. For completeness, we also include a summary of the peculiar changes of V79.

\begin{figure}
\includegraphics[width=8.2cm]{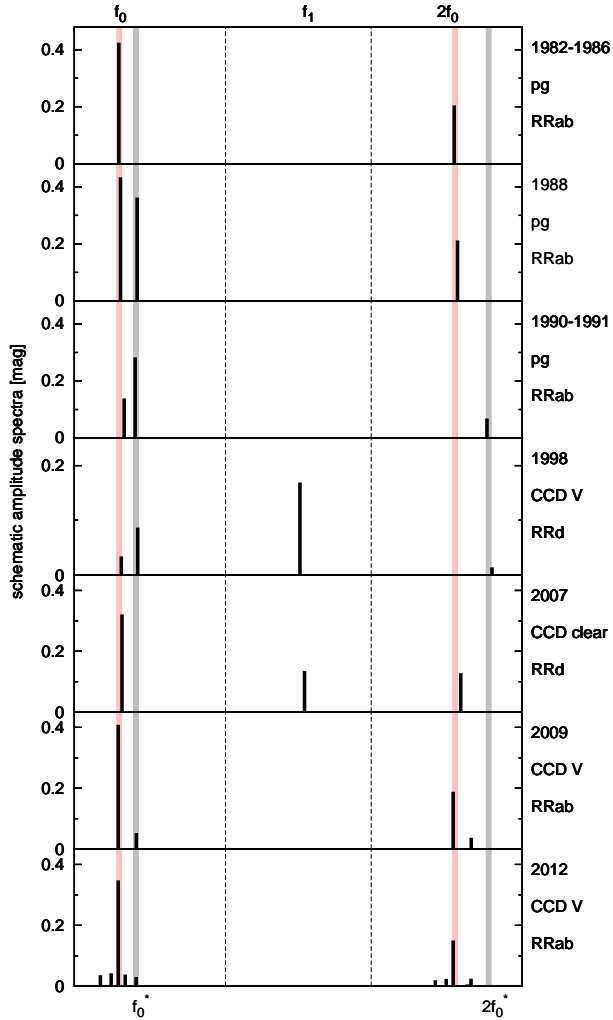}
\caption{The same as Fig.~\ref{13} for the frequency components of V79 in the $1982-2012$ period. Only those epochs are shown which give a substantially different solution.\label{79}}
\end{figure}

\begin{table}
\begin{center}
\caption{Period and magnitude changes of V79 between 1992 and 2012.\label{79mag.tab}}
\begin{tabular}{@{\hspace{0mm}}c@{\hspace{1mm}}l@{\hspace{1mm}}c@{\hspace{1mm}}c@{\hspace{2mm}}c@{\hspace{1mm}}c@{\hspace{2mm}}c@{\hspace{1mm}}c@{\hspace{2mm}}l}
\hline
\hline
Year &$P_{0}$&\multicolumn{2}{c}{$<B>$} &\multicolumn{2}{c}{$<V>$} &\multicolumn{2}{c}{$<I_{\mathrm C}>$}&Data\\
& &mag&int&mag&int&mag&int&\\
\hline
1992-1993& 0.47948&15.95& 15.94&15.67&15.66&&&C01\\
1996     & 0.4798 &      &      &15.71&15.70&& &K98\\
1998     & 0.47927& 16.00& 15.99&15.72&15.71&15.28&15.28 &H05,B06\\
1998     & 0.47927&      &      &15.71&15.70&& &CG\tablenotemark{a}\\
2009     & 0.48329&      &      &15.78&15.72&& &J12\\
2012     & 0.48329& 16.06& 16.00&15.76&15.72&15.31&15.29 &D12\\
\hline
\end{tabular}
\tablenotetext{1}{\cite{cl99}}
\end{center}
\end{table}

During the major part of its double-mode stage and just preceding it (between 1990 and 1998) the fundamental mode frequency was 0.016~d$^{-1}$ larger (the period 0.004~d shorter) than it was prior and after the double-mode episode. The amplitude of the overtone mode was about twice the amplitude of the fundamental mode in $1992-1998$ while the fundamental mode has became the dominant mode again from 2006. The frequency ratio of the radial modes was larger  in $1992-1998$ and smaller than normal in $2006-2007$ by about 0.002.

The changes in the frequency spectrum of V79 is documented in Fig.~\ref{79}. Only the substantially different solutions are shown.
Preceding the emergence of the overtone mode, the fundamental mode frequency had been splitted into two components and the amplitudes of the components had been transposed. Besides the dominant frequencies observed at 2.0711 and 2.0843~d$^{-1}$, residual signal appeared at 2.087 and 2.0743~d$^{-1}$ frequencies in the 1988 and $1990-1991$ photographic data, respectively. The combined 1998 CCD data (H05, B06, and CG99) showed the dominant fundamental frequency at 2.0865~d$^{-1}$ but a frequency component at 2.0719~d$^{-1}$ appeared in this data set, too. The frequency changes of the first harmonic frequency of $f_{\mathrm{F}}$ followed the displacement of the fundamental frequency.
After the double-mode episode,  several $kf_{\mathrm{F}}+f_{\mathrm m}\,\, (f_0=2.069,\,\,f_{\mathrm m}=0.016$~d$^{-1})$ components appeared in the 2009 data. The frequency of the strongest modulation component equals with the fundamental-mode frequency (2.085~d$^{-1}$) observed during the double-mode phase. Later, the modulation has become more complex, multiple periodic with 0.006 and 0.016~d$^{-1}$  frequencies of the primary and secondary modulations.  The  2.085~d$^{-1}$ frequency component appears in the D12 data, too. However,  it is only the fourth largest modulation component with 0.031 mag amplitude. 

Table~\ref{79mag.tab} lists the fundamental-mode period and the magnitude- and intensity-averaged mean $B,V, I_{\mathrm C}$ magnitudes of V79 derived from the CCD data. According to the pulsation equation \citep[given e.g. by][]{marc} a 0.0043 increase of log$L$, i.e., a 0.011~mag decrease in magnitude, accounts for a 0.004~d increase of the pulsation period. Though the changes of the intensity-averaged mean magnitudes may support that luminosity changes are standing behind the period increase of V79, we have to note also that differences of the order of 0.01~mag in  photometric results of GCs are common, simple because of  calibration and/or instrumental effects. Moreover, the much larger decrease of the magnitude averaged brightness may be the consequence of that  sinusoidal overtone-mode was dominant in $1992-1998$ but the large-amplitude, asymmetric RRab-type light curve observed in 2009 and 2012 is strongly nonlinear. Consequently, the observed increase of the intensity-averaged mean magnitudes may also arise, at least partly, from the differences of the light-curve shapes. \\
{\bf{V87}} All the CCD and photographic data yield a very similar solution with a dominant overtone mode. Period changes of both modes are marginal.
\\
{\bf{V99}} The period ratio of the radial modes are anomalously large according to each observation that shows double-mode behavior.  The reanalysis of the photographic data has revealed that besides the overtone mode the fundamental mode is also detectable in the majority of the data. Most probably, the improper phase coverage of the very-long-period Blazhko cycle of the fundamental mode (Sect.~\ref{blnotesec}) explains the previously detected changes of the fundamental-mode frequency. The $\sim$$0.001$~d$^{-1}$ decrease of the overtone-mode frequency is probably real. 
Although the frequency of the $f_{\mathrm{F}}+f_{\mathrm m}$ component detected in the D12 data is very close to the exact 3:4 resonance position to the overtone mode ($(f_{\mathrm{F}}+f_{\mathrm m})/f_1=0.7498$), the amplitude of this modulation component is only the half of the amplitude of $f_{\mathrm{F}}-f_{\mathrm m}$. Even in such close-resonance position, there is no sign that the 3:4 resonance would be in action \citep[see also in ][]{rrdbl}.\\
{\bf{V125}} The fundamental mode is detected only in the D12 data. Its very small amplitude makes the detection of this component in the previous data unlikely. The period of the overtone mode is very stable. The star is not over-luminous and lies at the blue edge of the double-mode region. 
\\
{\bf{V166}} Double-mode behavior is observed in each CCD data set. Both the fundamental- and the overtone-mode light curves are modulated. The period ratio is anomalously small.\\
{\bf{V200}} Each CCD data show double-mode behavior. The H05, B06, and the D12 data yield a similar light-curve solution with similar amplitudes of the fundamental and the overtone modes and with a normal period ratio. The C01 observations of V200 are not publicly available. Based on the analysis of a reprocessed version of the C01 data, C04  listed V200 as one of the double-mode stars with a too small period ratio. However, the strong aliasing of the C01 data makes the frequency determinations of C04 dubious. The strongest alias frequencies appearing in the C01 observations are separated by 0.003, 0.006, and 0.009~d$^{-1}$. The differences between the $f_{\mathrm{F}}$ and $f_{\mathrm{1O}}$ frequencies given by C04 and found in the other observations are $-0.006$ and 0.009~d$^{-1}$, which supports that, probably, no significant period change  occurs in this star. No difference between the amplitude ratios of the fundamental and first-overtone modes larger than expected from the uncertainties is seen between the solutions of the  H05, B06, and the D12 data. The amplitude changes of the modes found by C04 between the 1992 and 1993+1997 seasons are probably the consequence of the inadequacy of the data sampling and the mistaken frequency identification.  Based on the characteristics of its light curve  and period ratio, this star seems to be the twin of V68.\\
{\bf{V251}} The  frequencies published by C04 and B06 are significantly different from the results obtained from the D12 data. The reprocessed  C01 data of V251 analyzed in C04  are not publicly available. The differences between the fundamental- and overtone-mode frequencies  given by C04 and found in the present analysis are 0.009 and $-0.003$~d$^{-1}$. ~Thus,  it seems that aliasing might have biased the frequency determination  of V251 in C04, too. The frequencies  of V251 given by B06 are $0.03-0.04$~d$^{-1}$ larger than detected in the D12 data. Although the frequencies identified by B06 are indeed the largest amplitude signals in the Fourier spectrum of the data,  a solution with frequency values close to the results of the D12 data fits the observations with 7\% smaller residual scatter.  C04 reported significant amplitude changes of the modes, however, the D12 data do not show any frequency in the residual spectrum at the radial modes indicating that the light curves of the modes are stable.   \\
{\bf{V252}} The C01 observations of V252 are not publicly available.  The combined H05 and B06 data yield a similar solution as the D12 data. \\

\begin{figure*}
\includegraphics[width=15.4 cm]{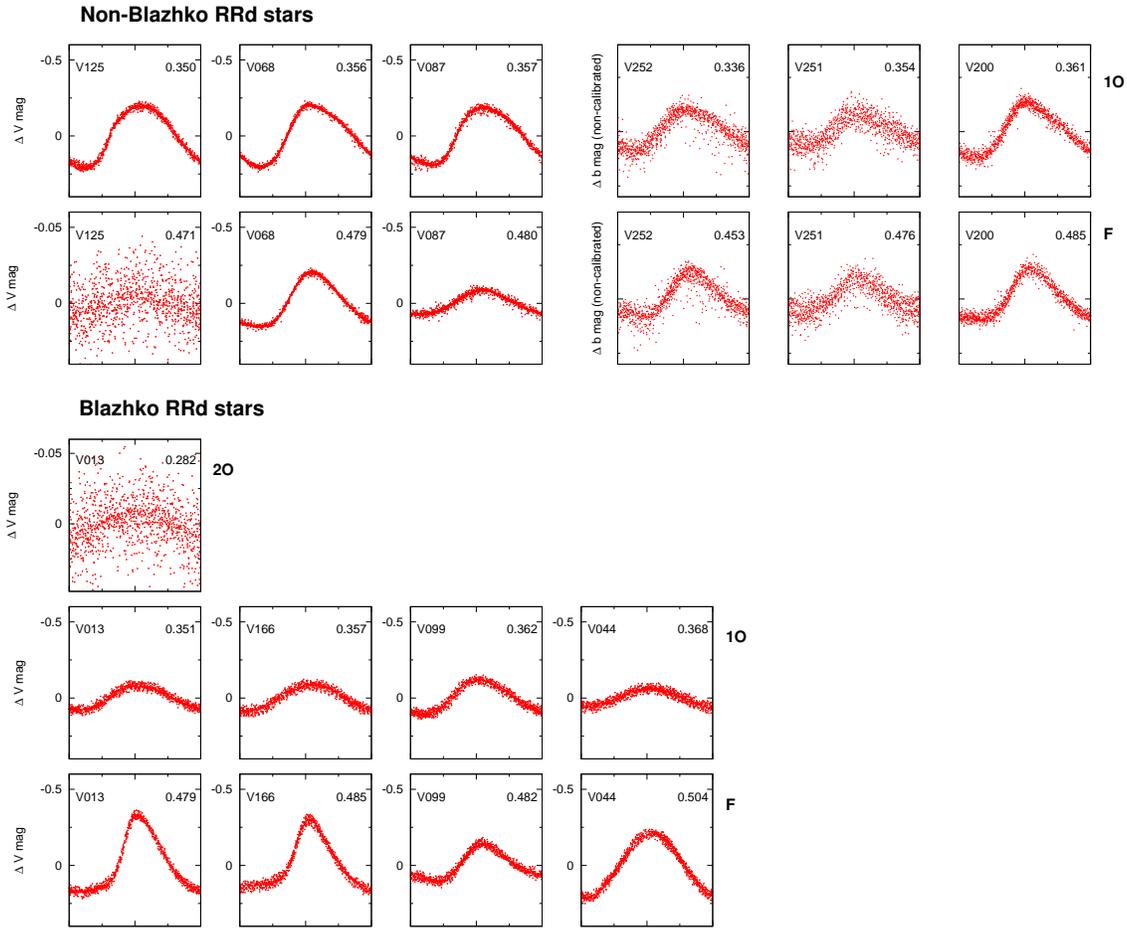}
\caption{Phased light curves of the radial modes of Blazhko and non-Blazhko double-mode stars. The periods (d) are given in the top-right-side of the panels. The data are prewhitened for all the other signals.\label{lcd}}
\end{figure*}

\begin{figure*}
\includegraphics[width=18.5 cm]{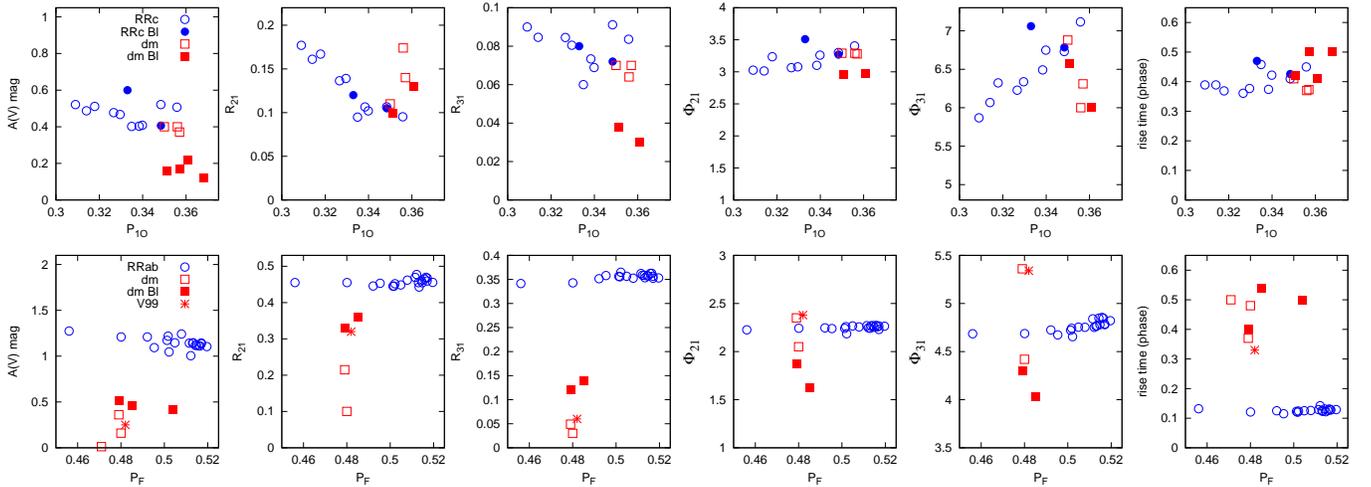}
\caption{The $V$ light-curve parameters (peak-top-peak $V$ amplitude, Fourier amplitude ratios and phase differences, and the rise time) of the first-overtone (top panels) and the fundamental (bottom panels) modes   of double-mode stars and of similar-period single-mode RRc and RRab stars are compared. The fundamental-mode parameters of V99 are denoted by a different symbol, because these parameters might not be reliable as only about half of the modulation cycle of the fundamental mode is covered by the observations. \label{fff}}
\end{figure*}

Our analysis shows that in normal-period-ratio RRd stars of M3  (V68, V87, V125, V200, V251, and V252) either the overtone mode is dominant  or the fundamental mode has a similar amplitude as the overtone. The majority of the known RRd stars share this property. Based on the results given in Table~\ref{dm.tab}, the modal content, i.e., the frequencies, and the amplitude- and period-ratios of V68 and V87 have been stable during the century covered by the observations. Most probably, V125, at the blue border of the double-mode region, is in the stage of the fall-off of the weak fundamental-mode component. This is supported by the similarity of the periods, amplitudes, colors and the light curves' shapes of V125 and V126 (the longest-period not over-luminous RRc star in M3). We suppose that the differences between the amplitude ratios and the frequencies published in C01, C04, H05 and B06 and obtained from the D12 data for V200, V251, and V252 do not reflect real changes of the double-mode properties of these stars, but arise from the strong aliasing problems of the previous data, which may have biased the previous frequency determinations. We conclude that  the pulsation properties of double-mode stars with normal period ratio are stable, they do not change significantly on a  timescale of about a century in M3.

\section{Discussion on the properties of Blazhko RRd stars}\label{rrdbsec}

RR Lyrae stars showing Blazhko effect  were first reported by \cite{so14}. A detailed analysis of these special variables found in the OGLE Galactic bulge collection were published recently by \cite{smo15a}. It is also probable that long-period  Blazhko modulation accounts for the detected changes of the amplitude ratios of the radial modes of NSVS 5222076 \citep{nsv}. The Blazhko modulation of 1O/2O double-mode Cepheids were discussed in detail by \cite{mk09}. 

Modulation is observed among 1O/2O LMC Cepheids either in both modes or neither of them  \citep{mk09}.  The same is true for F/1O double-mode RR Lyrae stars in M3. However, despite of the large amplitude modulations of the F and 1O modes, the small-amplitude 2O mode of V13 is not modulated.

\cite{mk09} explained the anti-phase modulations of the radial modes of Cepheids  by non-stationary resonant coupling of one radial mode with the another radial or a non-radial mode,  and with the sharing on the excitation sources between the two radial modes. Resonances are in fact detected in Cepheids, e.g. a 2:1 resonance produces well-seen features of the light curve of bump-Cepheids. Therefore, we have checked whether the light curves  of the radial modes of the Blazhko double-mode stars show any systematic difference from the light curves of other double-mode stars and from single-mode RR Lyrae stars. Fig.~\ref{lcd} shows the  light curves of the radial-modes of double-mode stars. The data were prewhitened  for the Fourier components of the other mode, for the modulation of both modes and for the linear-combination terms. The only conspicuous difference between the light curves of the Blazhko and non-Blazhko double-mode stars is that the overtone mode is the dominant mode (or the modes have similar amplitudes: V200, V251, V252) in RRd stars without light-curve modulation, while the fundamental mode has larger amplitude than the overtone in Blazhko RRd stars. 

 There is no well seen distortion in the mean light curves of the radial modes of  Blazhko stars (see in Fig.~\ref{lcd}). The fundamental-mode light curves of both Blazhko and non-Blazhko RRd stars differ  from the light curves of single-mode RRab stars systematically, however, the lack of the higher-order harmonic components in RRd stars may explain the differences mostly.

None of the light curves at different Blazhko phases shown in Fig.~\ref{blfig} indicates that any resonance would affect their shapes either. 
Amplitude and phase modulations of both modes of the  Blazhko double-mode stars are evident, however the light curve is smooth in each Blazhko phase. Only the fundamental mode of V44 displays irregularities, but this is produced by the complexity of the modulation probably.

The $V$ light-curve parameters of the fundamental and the first-overtone modes of double-mode stars and of similar-period single-mode stars are shown in Fig.~\ref{fff}. The period, peak-to-peak $V$ amplitude,  the $R_{21}$, $R_{31}$, $\varphi_{21}$,  $\varphi_{31}$ Fourier amplitude-ratios and phase-differences and the rise-time (in phase units) are plotted versus the overtone- and fundamental-mode periods in the panels. The  Fourier amplitude-ratios and phase-differences are not shown for the most sinusoidal components of the  double-mode stars. The only  systematic difference between the light-curve parameters of the overtone-mode of Blazhko and non-Blazhko RRd stars is that the $A$ amplitudes and the $R_{31}$ amplitude-ratios and maybe the  $\varphi_{21}$ phase differences of  Blazhko RRd stars are systematically smaller than those of normal RRd and RRc stars. 
The rise time of each RRd star is around the same as the rise time of RRc stars (between 0.4 and 0.5).
The amplitudes and amplitude-ratios of the fundamental mode of both Blazhko and non-Blazhko RRd stars are significantly smaller than those in normal RRab stars (the amplitude ratios of normal RRd stars are even smaller than of Blazhko RRd stars); the phase-differences of Blazhko RRd stars are also too small, while the phase-differences of normal RRd stars seem to  be scattered around the values of RRab stars. The rise time of the fundamental mode of each RRd star is significantly larger than the rise time of RRab stars. 

Based on the light curve's shapes and  the Fourier parameters of the Blazhko RRd stars we have not found any clear evidence   of a distortion caused by resonance interaction.

\begin{figure}
\includegraphics[width=8.4 cm]{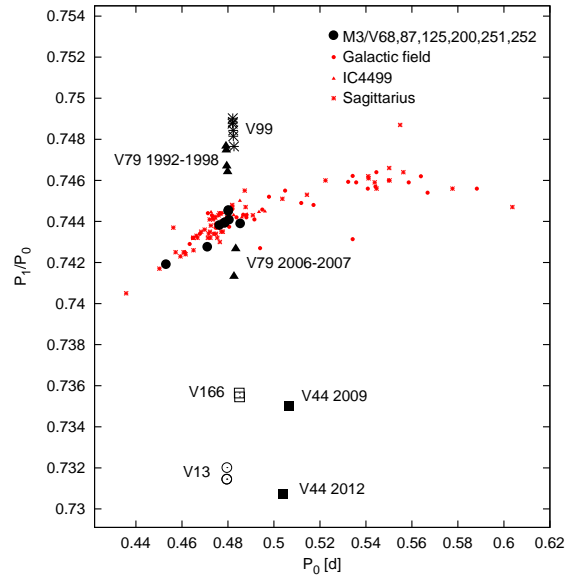}
\caption{Petersen diagram, $P_1/P_0$ period ratio vs $P_0$, of M3 double-mode stars is shown. Filled circles denote RRd stars with normal period ratio (V68, V87, V125, V200, V251, and V252). Other symbols denote the results obtained for Blazhko RRd stars at the different epochs according to the data given in Table~\ref{dm.tab}.  For comparison, the period ratios of double-mode stars in the Galactic field, IC 4499 and the Sagittarius dwarf galaxy are also shown.\label{dm.fig} }
\end{figure}

\begin{figure}
\includegraphics[width=8.4 cm]{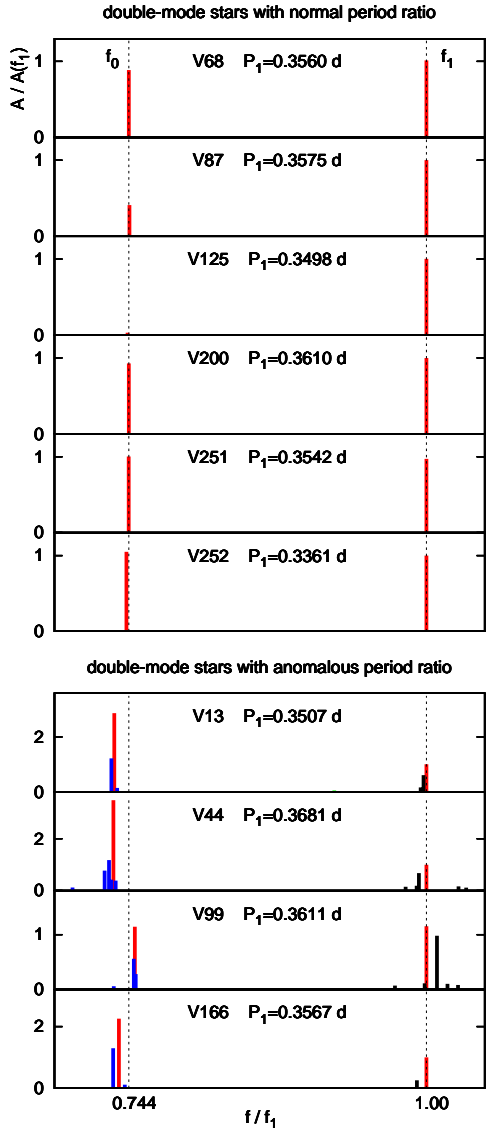}
\caption{The fundamental- and overtone-mode frequencies of RRd stars in M3. The radial-mode frequencies of RRd stars with normal period ratio ($P_1/P_0=0.742-0.745$) are separate, single signals. On the contrary, complex multiplets at the radial-mode frequencies, characteristics of Blazhko modulation, are detected in RRd stars, which have anomalous period ratios.\label{fr}}
\end{figure}

In \cite{rrdbl} we have shown that the anomalous period ratio is a common property of RRd stars showing Blazhko modulation of the radial modes in M3.
 The opposite is also true, none of the RRd stars with normal period ratio shows modulation of any of the modes. A similar tendency has been detected also in the Galactic bulge data \citep{smo15a}. The anomalous period ratios of RRd stars cannot be interpreted by any canonical explanation  as it was discussed in detail in C04 and in \cite{smo15a}.

In Fig.~\ref{dm.fig}, the $P_{\mathrm{1O}}/P_{\mathrm{F}}$ period ratio of M3 double-mode stars are shown in comparison with results of the double-mode stars in the Galactic field, the IC 4499 globular cluster and the Sagittarius dwarf galaxy \citep[compilation by][]{mcc,wils,wn,cs}. The period ratios of six RRd stars (V68, V87, V125, V200, V251, and V252) fit the region covered by the data of the other systems. The period ratios of Blazhko RRd stars (V13, V44, V79, V99, and V166) are  anomalous according to each of the data, which show double-mode behavior (see Table~\ref{dm.tab}). 

The frequencies appearing at the positions of $f_{\mathrm{F}}$ and $f_{\mathrm{1O}}$ of RRd stars in the D12 data are shown in Fig.~\ref{fr}. Contrary to the single signals of the radial modes of normal period-ratio RRd stars, the frequency spectrum of Blazhko RRd stars are comprised from multiplets  at the fundamental- and the overtone-mode frequencies. The highest-amplitude frequency belongs to the fundamental mode in these stars. 

The data given in Table~\ref{dm.tab} and Figs.~\ref{13}, \ref{44} and ~\ref{79} indicate that in Blazhko RRd stars not only the period ratios are anomalous, but from time to time, a different component of the multiplets is the dominant component. OGLE-BLG-RRLYR-02530 \citep{smo15a} seems to be a similar case. We have also checked whether any pair of the components of the multiplets  would yield a period ratio in the required, $0.743-0.744$ range.
However, the mystery of the anomalous period ratios of Blazhko RRd stars cannot be resolved in this way; none of the components of the multiplets can give a proper period ratio.


Connected  to  the switch  between the single- and double-mode states (just preceding or following it), as large as $0.010-0.015$~d$^{-1}$ jumps of the fundamental-mode frequency ($0.003-0.004$~d jumps in the periods) are detected in the Blazhko RRd stars, V13, V44, and V79. The frequency shift preceding the switch to double-mode stage is always positive (the period is decreasing) and it is accompanied by a strong amplitude decrease of the fundamental mode in all these stars.

Blazhko RRab stars are  known to show  large, irregular, rapid period changes (J12), however, all these changes are at least about an order of a magnitude smaller than the period changes observed in Blazhko double-mode stars.
The switch from double-mode to fundamental-mode pulsation of OGLE-BLG-RRLYR-12245 \citep{so14} was accompanied by a  0.0005 d period increase ($-0.003$  d$^{-1}$ in frequency). The period change was attributed to a nonlinear shift caused by the four-times increase of the amplitude of the fundamental mode by \cite{so14}.   
Although  simultaneously with the  appearance of the overtone mode, amplitude decrease of the  fundamental-mode of V13, V44, and V79 are also detected, as their period shifts are about an order of a magnitude larger than in OGLE-BLG-RRLYR-12245, therefore, nonlinearity cannot explain the observed frequency shifts in these stars completely.

The displacement of the frequencies corresponds to the modulation frequency (half of it in V13) indicating that, instead of  real frequency changes, the modulation is so strong  that the side frequency has the largest amplitude in these cases.
The dominant component is the $f+f_{\mathrm m}$ frequency  in the double-mode stage of these three stars. 
Thus, the anomalously low period ratios would become even lower if the pre-RRd frequency of the fundamental-mode were taken as the real fundamental-mode frequency in V13 and V44.

Anomalous period ratios are detected mostly (probably exclusively) in Blazhko RRd stars. Assuming that pulsation models give correct results for the period ratios of RR Lyrae stars (as in the case for non-Blazhko RRd stars) we have to think that the stellar structure of Blazhko and non-Blazhko stars should be different in order to explain the anomalous period ratios of Blazhko RRd stars. 

\begin{figure}
\includegraphics[width=8.9 cm]{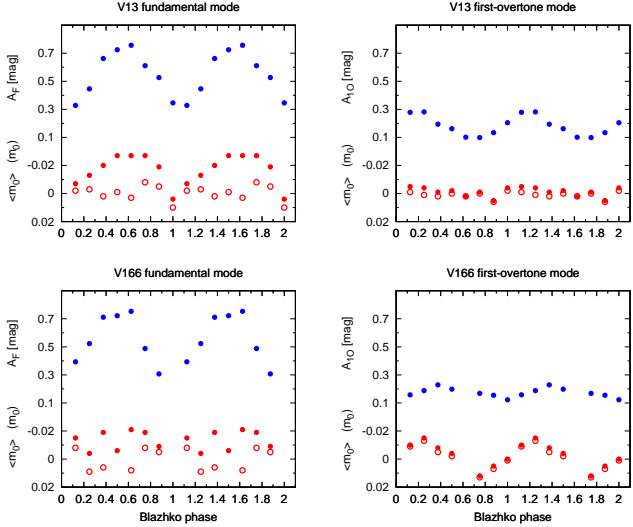}
\caption{Variations of the fundamental- (left-side panels) and overtone-mode (right-side panels) amplitudes and the magnitude-  and intensity-averaged  mean $V$ magnitudes ($<m_0>$, and $(m_0)$ shown by open and filled circles, respectively) of the light curves of V13 (top panels) and V166 (bottom panels) during the Blazhko cycle. Note that the Blazhko periods of the fundamental and the overtone modes are the same for V13, but they are respectively 71.5~d and 43.5~d for V166. The results are derived for the light curves in the eight Blazhko phases shown in Fig.~\ref{blfig}. The light curves are prewhitened for the pulsation and modulation of the other radial mode and for the linear-combination terms. 
\label{a-m.fig}}
\end{figure}

\begin{figure}
\includegraphics[width=8.6 cm]{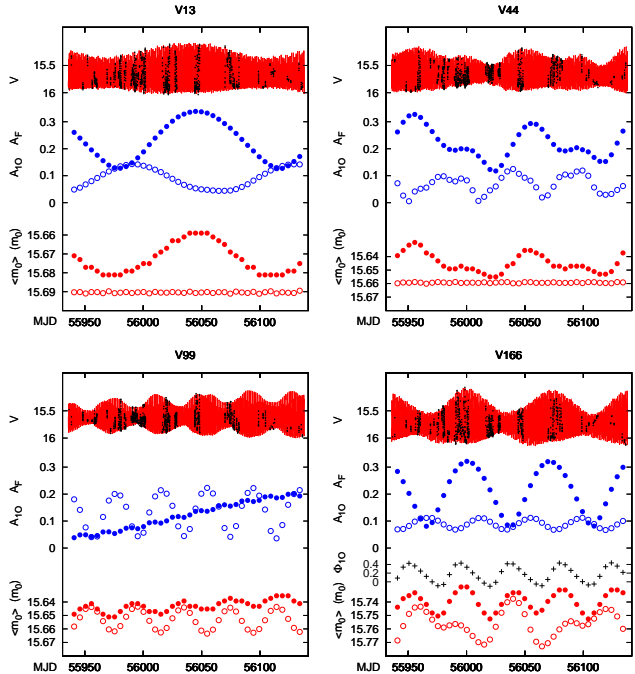}
\caption{Amplitudes ($A_{\mathrm {1O}}$ and $A_{\mathrm F}$ shown by open and filled blue circles, resp.) and magnitude- and intensity-averaged mean-brightness (shown by open and filled red circles, resp.) of Blazhko RRd stars derived from 5-d segments of the synthetic data generated from the full light-curve solutions.  In the top part of each plot the synthetic light curves are shown with the observations over-plotted. The phase variation of the overtone mode of V166 is also shown by `+' symbols.\label{ereduj.fig}}
\end{figure}

The  period ratios of the modulations of the overtone and the fundamental modes ($Pm_{\mathrm{1O}}/Pm_{\mathrm{F}}$) of the double-mode Blazhko stars (V13, V44,  V99, and V166)  match the 1:1, 4:7, 1:10, and 3:5 small-integer ratios within about the error limits of the periods. The modulation-period ratios of some of the Galactic bulge Blazhko RRd stars are also close to small integer ratios (e.g. it is close to 3:10, 2:3, 1:3, 4:7, 1:5, 2:3 in OGLE-BLG-RRLYR-04598, -07393, -10728, -11311, -22356, and -32462, respectively). 

It was already shown that the light curves of some Blazhko RRab stars were modulated with different periods either simultaneously or at different epochs  and that the ratios of the modulation-periods of multiple-periodic Blazhko stars were close to the ratios of small integer numbers \citep{czl,m5bl,rzl,kepler}. In other stars, different components of the multiplets reflect the dominant modulation at different epochs, i.e., the amplitudes of the multiplet components vary in time, according to the long-term data as e.g. in V44,V140/M3 (this paper)  and V2/M5 \citep{m5bl}. In \cite{m5bl}, we wrote `If these modulations correspond to elements of series of equidistant spacing multiplets, their frequency ratios are close to fractions of small-integer numbers. This indicates that the observed modulation frequency values may not always correspond to the `base frequency', i.e., the shortest frequency component of a multiplet'. The ratios of the modulation periods  of the F and 1O modes in Blazhko double-mode stars may be explained similarly. If both modes select their modulation frequency from the same, rich multiplet spectrum,  than any small-integer ratio of the modulation periods is equally possible.

\cite{smo15a} found a `feedback effect' between the star's brightness and pulsation-mode amplitude in some of the OGLE Blazhko RRd stars. They noticed that: `larger the amplitude of the dominant mode, smaller the mean stellar brightness'. 

The connection between the mean brightness and amplitude variations is checked both on the observed and on synthetic data. The  magnitude- and intensity-mean $V$ brightnesses and the amplitudes of the F and 1O modes  are shown in Fig.~\ref{a-m.fig} as derived from the  disentangled light curves of the radial modes of V13 and V44, the two Blazhko RRd stars which have good pulsation-phase coverage of both radial modes in most of the Blazhko phases. The same parameters are also determined by fitting the radial-mode frequencies and their linear combinations
to 5-d segments of synthetic data of the 4 Blazhko RRd stars (Fig.~\ref{ereduj.fig}). The synthetic data are generated from the modulated radial modes' components of the full light-curve solutions.

 Mean-brightness variations are indeed detected in each of the four Blazhko RRd stars.
The intensity-averaged mean brightness differ significantly from the magnitude-averaged value in some cases, because of the non-sinusoidal shape of the light curves, which is especially true for the fundamental mode in the large-amplitude phase of the modulation. If we accept that the intensity-averaged mean-brightness variation reflects the luminosity changes, as it was shown in Blazhko stars \citep[e.g. in][]{mw2,czl,rzl} using the photometric Baade-Wesselink method \citep{ip}, the brightness variations of V13 and V44 are in phase with the amplitude variations of the dominant fundamental mode. The intensity-averaged mean-brightness variation of V99 is twofold, it varies in anti-phase with the amplitude changes of the overtone mode, and, meanwhile, a gradual brightness increase, parallel with the amplitude increase of the fundamental mode, is also observed. The amplitude changes of V166 are the most surprising: the intensity-averaged mean brightness does not show any systematic changes with the modulation period of the dominant fundamental mode, which shows large-amplitude modulation, instead, it varies with the modulation period of the overtone mode. The brightness variation is in a $\pi/2$ phase lag with the very small-amplitude changes of the overtone mode.  Actually, the intensity-averaged mean brightness  follows the phase variation (also shown in Fig.~\ref{ereduj.fig} for V166) of the overtone mode. As a summary, no universal connection between the brightness and the amplitude variations seems to exist.

 
\section{Summary}\label{sumsec}
The light variation of overtone and double-mode RR Lyrae stars has been investigated based on an extended new $BVI_{\mathrm C}$ time-series photometry and on archive data. The GC data has a unique potential to map the distribution of variables showing different properties and extra features in a large and homogeneous sample of stars.

Peculiarities of the behavior  of stars showing a frequency component at 0.61 frequency ratio, the Blazhko effect, period doubling, and double-mode pulsation are discussed in all details.

The most important findings of our study are summarized in the followings.

\begin{enumerate}

\item{The $f_{0.61}$ frequency is observed in 14 RRc stars, three normal and one anomalous period-ratio RRd stars. Sect.~\ref{f6sec}.}

\item{The amplitude and frequency of the $f_{0.61}$  component show considerable changes, while the radial mode component remains stable. Sect.~\ref{f6stabsec}.}

\item{The light curves of RR$_{0.61}$ stars are sinusoidal, with a reduced, if any, bump preceding maximum brightness. Sect.~\ref{f6lcsec}.}

\item{The $f_{0.61}$ frequency occurs in 4 RRd stars, in each RRc star at the blue-side of the double-mode region, and in some bright, evolved variables with periods longer than the period of the 1O-mode of RRd stars. Actually, it is detected in each RRc star with $B-V \gtrsim 0.24$~mag and $V-I \gtrsim 0.36$~mag color indices. Connection between the double-mode pulsation and the appearance of the $f_{0.61}$ frequency is proposed based on the distribution  RR$_{0.61}$ stars. Sect.~\ref{f6parsec}.}

\item{Period doubling of the small-amplitude 2O mode of V13  ($3f_{\mathrm{2O}}/2$) and the fundamental mode of  V166 ($7f_{\mathrm F}/2$) are detected. Blazhko effect of both the F and 1O modes of V13 and V166 are observed, but the 2O mode of V13 does not show any modulation. No half-integer component of any first-overtone or $f_{0.61}$ frequency has been detected.  Sect.~\ref{pdsec}.} 

\item{Blazhko modulation of both radial modes of four RRd stars, and three luminous  RRc stars are observed. A small-amplitude frequency component appears in the residual spectra of one normal-brightness RRc star and in a 2O variable close to the overtone-mode frequency. Strong phase modulation characterizes the modulation of the overtone modes. Sect.~\ref{blsec}.}

\item{Two new double-mode stars are identified: F-mode frequency is detected in the RRc star, V125, and 1O-mode frequency in the RRab star, V44. Sect.~\ref{dmsec}.}

\item{The most complex variability is detected in V13, which shows three radial modes, strong Blazhko effect of the F and 1O modes, period doubling of the 2O mode, and the $f_{0.61}$ frequency component, as well.  Although modeling the radial-mode frequencies of triple-mode stars may yield accurate physical parameters, unfortunately, this is not the case in V13. Pulsation models fail to explain the anomalous period ratios of some RRd stars observed in M3 and in the Galactic bulge \citep[C04,][]{smo15a}, and V13 is one of the Blazhko RRd stars with anomalously small $P_{\mathrm{1O}}/P_{\mathrm{F}}$ period ratio. Sect.~\ref{v13sec}.}

\item{The modal content and the frequencies of RRd stars seems to be remaining stable during the timescale of the photometric observations of M3 for variables with period ratios matching  the sequence defined by the majority of the known RRd stars on the $P_{\mathrm{1O}}/P_{\mathrm{F}}$-$P_{\mathrm{F}}$ Petersen diagram (V68, V97, V200, V251, and V252). Sect.~\ref{dmnotesec}.} 

\item{ The period-ratio of four RRd stars  (V13, V44, V99, and V166) are anomalous, and it was the case for V79 during its double-mode stage, as well. To explain the anomalous period ratios of Blazhko RRd stars is a great challenge for stellar pulsation theory. The anomalous period ratios of Blazhko RRd stars can,  in principle, be explained by extreme high- or low-metallicity models \citep[see fig. 7. in][]{smo15b}, however, it is unlikely in about 50\% of the RRd stars in M3, as no extreme metallicity value of any other star of the cluster has been reported. Each anomalous period-ratio RRd stars  shows Blazhko modulation of both the F and 1O radial modes. The fundamental mode is dominant in these star.  Sects.~\ref{dmnotesec},~\ref{rrdbsec}.} 

\item{No evidence of a resonance-induced systematic deformation of the light curves of Blazhko RRd stars is found, still in the case of V13, where the modulations of the F and 1O modes have the same periods and they show anti-phase amplitude changes similarly to 1O/2O Cepheids. Sect.~\ref{rrdbsec}.}

\item{The M3 data show, that the `feedback effect' \citep{smo15a} between the mean brightness and the amplitude of the dominant mode in Blazhko RRd stars does not exist in general. 
The connection between the amplitude variations of the modes and the mean brightness (defined as the intensity-averaged mean $V$ magnitude) is  different star by star. Sect.~\ref{rrdbsec}.}

\item{Large period jumps are recorded in V13, V44, and V79, which are connected to the switch from fundamental- to double-mode pulsation.  The value of the frequency increase  equals with the modulation frequency (or its half), and  strong amplitude decrease accompanies the appearance of the overtone mode in Blazhko RRab stars.  Different components of the Blazhko multiplets seem to be the highest amplitude signals at the fundamental  mode frequency in single- and double-mode stages in these stars. Sects.~\ref{dmnotesec},~\ref{rrdbsec}.}

\item{The period-ratio anomaly of Blazhko RRd stars cannot be resolved with any pair of the components of the multiplets appearing at the radial modes. Sect.~\ref{rrdbsec}.}

\end{enumerate}

As a conclusion we would like to put into the focus the strange $f_{0.61}$ signals detected in different samples of classical radial pulsating stars and the special anomalous period ratios of RRd stars observed in this M3 survey as open questions requiring further investigation. 

\section*{Acknowledgments}
The authors would like to express their acknowledgement to the referee for the notes, questions and comments, which helped us to improve the clarity and quality of the paper.
GH acknowledges support by the Ministry for the Economy, Development, and Tourism’s ssyPrograma Iniciativa Cient\'ifica Milenio through grant IC210009, awarded to the Millennium Institute of Astrophysics; by Proyecto Basal PFB-06/2007; by Fondecyt grant 1141141; and by CONICYT Anillo grant ACT1101.   KK is supported by a Marie Curie International Outgoing Fellowship within the 7th European Community Framework Programme (PIOF 255267  SAS-RRL).
NJ acknowledges the OTKA K-83790 grant, MA the grant of the MTA CSFK Lend\"ulet Disk Research Group, AP the LP2012-31, K-109276 and K-104607 grants. KS has been supported by the Lend\"ulet-2009 program of the HAS, the Hungarian OTKA grant K-104607 and the ESA PECS Contract No. 4000110889/14/NL/NDe.


\begin{thebibliography}{99}
\bibitem[Alard(2000)]{isis} Alard, C. 2000, A\&AS, 144, 363
\bibitem[Benk\H{o} et al.(2006)]{be06} Benk\H{o}, J. M., Bakos, G. \'A., \& Nuspl, J. 2006, MNRAS, 372, 1657 (B06)
\bibitem[Benk\H{o} et al.(2014)]{kepler} Benk\H{o}, J. M., Plachy, E., Szab\'o, R., Moln\'ar, L., \& Koll\'ath, Z. 2014, ApJS, 213, 31
\bibitem[Buchler \& Koll\'ath (2011)]{kb} Buchler, R., \&  Koll\'ath, Z. 2011, ApJ, 731, 24
\bibitem[Castelli \& Kurucz(2003)]{kurucz} Castelli, F. \& Kurucz, R. L. 2003, in Modelling of Stellar Atmospheres, ed. N. Piskunov, W.-W. Weiss and D.-F. Gray, IAU Symp., 210, 20
\bibitem[Cacciari et al.(2005)]{ca05} Cacciari, C., Corwin, T.M., \&  Carney, B.W. 2005, AJ, 129, 267
\bibitem[Clement \& Goranskij(1999)]{cl99} Clement, C. M., \&  Goranskij, V. P. 1999, ApJ, 513, 767
\bibitem[Clement et al.(1997)]{cl97} Clement, C. M., Hilditch, R. D., Kaluzny, J., \&   Rucinski, S. M. 1997, ApJ, 489, L55 
\bibitem[Clement et al.(2001)]{cl01} Clement, C. M., Muzzin, A., Dufton, Q., Ponnampalam, T., Wnag, J., Burford, J., Richardson, A., \&  Rosebery, T. 2001 AJ, 122, 2587
\bibitem[Clement \& Thompson(2007)]{v79ct} Clement, C. M., \&   Thompson, M. 2007, JAVSO, Vol 35., no. 2, p. 336
\bibitem[Clementini et al.(2004)]{cc04} Clementini, G., Corwin, T. M., Carney, B. W., \&   Sumerel, A.N. 2004, AJ, 127, 938 (C04)
\bibitem[Corwin \& Carney(2001)]{co01} Corwin, B. W., \&   Carney, B. W. 2001, AJ, 122, 3183 (C01)
\bibitem[Cseresnjes(2001)]{cs} Cseresnjes, P. 2001, A\&A, 375, 909
\bibitem[Dziembowski(2012)]{dz12} Dziembowski, W. A. 2012, AcA, 62, 323
\bibitem[Garrido(2004)]{g04} Garrido, R., 2004, in Second Eddington Workshop: Stellar Structure and Habitable Planet Finding, ed. Favata F., Aigrain S., Wilson A., ESA SP-538, p. 231
\bibitem[Gillet \& Fokin(2014)]{gf} Gillet, D., \&   Fokin, A. B. 2014, A\&A, 565, 73
\bibitem[Goranskij et al.(2010)]{go10} Goranskij, V. P., Clement, C. M., \&   Thompson, M. 2010, in Variable Stars, the Galactic halo and Galaxy Formation, ed. Sterken C., Samus N. \& Szabados L. (Moscow: Sternberg Astronomical Institute of Moscow Univ.), p. 115
\bibitem[Goranskij \& Barsukova(2007)]{gb07} Goranskij, V. P.,  \&   Barsukova, E. A. 2007, Astron. Telegraph, No. 1120
\bibitem[Govea et al.(2014)]{ggp} Govea, J., Gomez, T., Preston, G. W., \&   Sneden, Ch. 2014, ApJ, 782, 59\bibitem[Hartman et al.(2005)]{H05} Hartman, J. D., Kaluzny, J., Szentgy\"orgyi, A., \&   Stanek K. Z. 2005, AJ, 129, 1596 (H05)
\bibitem[Hurdis \& Krajci(2012)]{nsv} Hurdis, D. A., \&   Krajci, T. 2012, JAAVSO, Vol. 40, 268
\bibitem[Jurcsik et al.(2008)]{mw2}  Jurcsik, J., S\'odor, \'A., Szeidl, B., et al. 2008, MNRAS, 393, 1553
\bibitem[Jurcsik(2009)]{aip} Jurcsik, J. 2009, in: Stellar pulsation: challenges for theory and observation, ed. J.A. Guzik and P.A. Bradley, AIP Conf. Proc. 1170, 286
\bibitem[Jurcsik et al.(2009)]{kbs1} Jurcsik, J.,S\'odor, \'A., Szeidl, B., et al. 2009, MNRAS, 400, 1006
\bibitem[Jurcsik et al.(2011)]{m5bl} Jurcsik, J., Szeidl, B., Clement, C., Hurta, Zs., \&   Lovas, M. 2011, MNRAS, 411, 1763
\bibitem[Jurcsik et al.(2012a)]{oc} Jurcsik, J., Hajdu, G., Szeidl, B., et al. 2012a, MNRAS, 419, 2173 (J12)
\bibitem[Jurcsik et al.(2012b)]{rzl} Jurcsik, J., S\'odor, \'A., Hajdu, G., et al. 2012b, MNRAS, 423, 993
\bibitem[Jurcsik et al.(2014)]{rrdbl} Jurcsik, J., Smitola, P., Hajdu, G., \&   Nuspl, J. 2014, ApJ Letters, 797, L3
\bibitem[Kaluzny et al.(1998)]{K98} Kaluzny J., Hilditch, R. W., Clement, C., \&   Rucinski, S. M. 1998, MNRAS, 296, 347 (K98)
\bibitem[Koll\'ath(1990)]{mufran} Koll\'ath, Z. 1990, Occ. Techn. Notes Konkoly Obs., No. 1, http://www.konkoly.hu/staff/kollath/mufran.html
\bibitem[Koll\'ath et al.(2011)]{k11} Koll\'ath, Z., Moln\'ar, L., \&   Szab\'o, R. 2011, MNRAS, 414, 1111
\bibitem[Le Borgne  et al.(2012)]{geos} Le Borgne, J.-F., Klotz, A., Poretti, E. et al. 2012, AJ, 144, 39
\bibitem[Marconi \& Degl'Innocenti(2007)]{marc} Marconi, M., \&   Degl'Innocenti, S. 2007, A\&A, 474, 557
\bibitem[McClusky(2008)]{mcc} McClusky, J. V. 2008, IBVS, 5825
\bibitem[Moskalik \& Ko\l aczkowski(2008)]{mk08} Moskalik, P., \&   Ko\l aczkowski, Z. 2008, Comm. in Astroseismology, 157, 343
\bibitem[Moskalik \& Ko\l aczkowski(2009)]{mk09} Moskalik, P., \&   Ko\l aczkowski, Z. 2009, MNRAS, 394, 1649
\bibitem[Moskalik (2014)]{mo14} Moskalik, P. 2014, in Precision Asteroseismology, ed., J.A. Guzik, W.J. Chaplin, G. Handler \& A. Pigulski, Proceedings IAU Symposium No. 301, 249
\bibitem[Moskalik et al.(2015)]{mo15} Moskalik, P., Smolec, R., Kolenberg, K., et al. 2015, MNRAS, 447, 2348
\bibitem[Munari et al(2005)]{munari} Munari, U., Sordo, R., Castelli, F., \&   Zwitter, T. 2005, A\&A, 442, 1127
\bibitem[Netzel et al.(2015)]{netzel} Netzel, H., Smolec, R., \&   Moskalik, P. 2015, MNRAS, 447, 1173
\bibitem[Peterson, Rood \& Crocker(1995)]{peterson} Peterson, R. C., Rood, R. T., \&   Crocker, D. A. 1995, ApJ, 453, 214
\bibitem[Preston \& Chadid(2013)]{pc}  Preston, G. W., \&   Chadid, M. 2013, EAS Publications Series, Volume 63, 35.
\bibitem[Samus et al.(2009)]{samus} Samus, N. N., Kazarovets, E. V., Pastukhova, E. N., Tsvetkova, T. M., Durlevich, O. V. 2009, PASP, 121, 1378
\bibitem[S\'odor et al.(2011)]{czl} S\'odor, \'A., Jurcsik, J., Szeidl, B., et al. 2011, MNRAS, 411, 1585
\bibitem[S\'odor et al.(2009)]{ip} S\'odor, \'A., Jurcsik, J., \&   Szeidl, B. 2009, MNRAS, 394, 261
\bibitem[S\'odor et al.(2012)]{nl} S\'odor, \'A. 2012 Occ. Techn. Notes Konkoly Obs., No. 1, http://www.konkoly.hu/staff/sodor/lcfit.html
\bibitem[Smolec (2015a)]{smo15a} Smolec, R., Soszy\'nski, I., Udalski, A., et al. 2015a, MNRAS, 447, 3756
\bibitem[Smolec (2015b)]{smo15b} Smolec, R., Soszy\'nski, I., Udalski, A., et al. 2015b, MNRAS, 447, 3873
\bibitem[Soszy\'nski et al.(2008)]{so08} Soszy\'nski, I., Poleski, R., Udalski, A., et al. 2008, AcA, 58, 163
\bibitem[Soszy\'nski et al.(2010)]{so10} Soszy\'nski, I., Poleski, R., Udalski, A., et al. 2010, AcA, 60, 17 
\bibitem[Soszy\'nski et al.(2011)]{so11} Soszy\'nski, I., Dziembowski, W. A., Udalski, A., et al. 2011, AcA, 61, 1 
\bibitem[Soszy\'nski et al.(2014)]{so14} Soszy\'nski, I., Dziembowski, W. A., Udalski, A., et al. 2014, AcA, 64, 1
\bibitem[Stetson(2000)]{st00} Stetson, P. B. 2000, PASP, 112, 925
\bibitem[Szab\'o(2014)]{sz14} Szab\'o, R., 2014, Precision Asteroseismology, Proceedings of the International Astronomical Union, IAU Symposium, Volume 301, pp. 241-248
\bibitem[Szab\'o et al.(2014)]{corot} Szab\'o, R., Benk\H{o}, J. M., Papar\'o, M., et al.  2014, A\&A, 570, 100
\bibitem[Szentgy\"orgyi  et al.(2011)]{mmt} Szentgy\"orgyi, A., F\H{u}r\'esz, G., Cheimets, P. et al. 2011, PASP, 123, 1188
\bibitem[Walker \& Nemec(1996)]{wn} Walker, A. R. \& Nemec, J. M. 1996, AJ, 112, 2026
\bibitem[Wils(2010)]{wils} Wils, P. 2010, IBVS No 5955
\end{thebibliography}
\end{document}